\documentclass[paper]{JFM-FLM_Au}
\usepackage{amsmath,amssymb}
\usepackage{graphicx}
\usepackage{dcolumn}
\usepackage{bm}
\usepackage{color}
\usepackage{tikz}
\usetikzlibrary{arrows.meta,positioning,calc}
\usepackage{pgfplots}
\usepackage{quiver}
\usepackage{mathtools}
\usepackage{array}
\usepackage{makecell}
\usepackage{cleveref}
\usepackage{tabularx}
\usepackage{enumitem}
\setlist[itemize]{noitemsep}

\pgfplotsset{compat=1.18}

\let\Delta\rmDelta

\title{An abundance of periodic orbits in two-dimensional turbulence revealed by a diffusion-based generative model}

\author{Jeremy P. Parker\aff{1}, Tobias M. Schneider\aff{2}}
\affiliation{
\aff{1}Mathematics, University of Dundee, Dundee DD1 4HN, United Kingdom
\aff{2}Emergent Complexity in Physical Systems Laboratory (ECPS), \'Ecole Polytechnique F\'ed\'erale de Lausanne, 1015 Lausanne, Switzerland
}

\corresau{Jeremy P. Parker, \email{jparker002@dundee.ac.uk}}

\lefttitle{J.P. Parker and T.M. Schneider}
\righttitle{Periodic orbits in turbulence revealed by a diffusion model}

\date{\today}

\begin{document}

\maketitle

\begin{abstract}
Unstable periodic orbits offer a route to reconstructing the statistical properties of chaotic systems, including  turbulence. Short periodic orbits are particularly important in periodic-orbit expansions, yet their abundance and diversity remain largely unexplored even in the extensively studied two-dimensional Kolmogorov flow. 
We search for short relative periodic orbits in this flow at $Re=40$, considering all distinct classes of recurrence symmetry.
We train a generative diffusion model on nonperiodic vorticity sequences from a direct numerical simulation of turbulence, producing time series that resemble physically plausible trajectories.
The model incorporates symmetry properties of the system but is otherwise trained solely on these data, without enforcing the governing equations. After training, we modify its architecture to generate synthetic trajectories that are periodic up to spatial symmetry transformations, without retraining its weights. These synthetic trajectories are then refined into exact  solutions using an iterative solver, yielding 111 new relative periodic orbits with periods $T < 3$, including 40 with $T<2$. Rather than revealing a hierarchy dominated by a small number of short cycles, this abundance suggests that quantitatively converged periodic-orbit descriptions of spatiotemporal chaos may require substantially larger orbit libraries than anticipated.
\end{abstract}

\section{Introduction}

The fluid dynamics of turbulence is the archetypal example of a high-dimensional chaotic system in which the governing equations are known, yet the dynamics they generate remain extraordinarily difficult to describe and interpret. One approach to addressing the complexity of dissipative chaos is based on comparatively simple, non-chaotic, invariant solutions of the governing equations, including equilibria, travelling waves, and periodic or relative periodic orbits. These exact solutions provide reproducible structures around which the geometry and dynamics of turbulence may be organized \citep{Kawahara2001periodic,Kawahara2012significance,Graham2021exact}.

Periodic orbits, including relative periodic orbits in systems with symmetries, are particularly important because they offer a connection between individual recurrent flow processes and the statistical properties of chaos. Each such orbit represents an exactly recurrent evolution, possibly up to a symmetry transformation, and can therefore encode a characteristic flow mechanism associated with turbulent dynamics. More ambitiously, periodic orbit theory (POT) expresses statistical averages over a chaotic attractor in terms of its unstable periodic orbits \citep{ChaosBook,cvitanovic1995dynamical}. Although the formal description involves infinitely many periodic orbits (or `cycles'), its practical appeal rests on the possibility of rapidly convergent expansions, such that a comparatively limited collection of periodic orbits already provides an accurate statistical and phenomenological description of the chaotic dynamics.

Short periodic orbits can play a particularly important role in cycle expansions. A symbolic dynamics describes trajectories by sequences of symbols labelling elementary passages through state space; periodic trajectories are represented by repeating words. In systems with a suitable symbolic organization, longer cycles closely shadow sequences of shorter ones. Their weighted contributions approximately cancel against products of shorter-cycle contributions, leaving small corrections and permitting rapidly convergent estimates of dynamical averages \citep{ChaosBook,cvitanovic1995dynamical,lan2010cycle}. The practicality of this construction depends on how many elementary cycles and admissible combinations must be represented, even at low symbolic order. Establishing how numerous the short cycles are is therefore an important step towards assessing the practicality of such descriptions.

Two-dimensional Kolmogorov flow provides a useful setting in which to investigate this question. At $Re=40$, the system exhibits sustained spatiotemporal chaos and has been extensively studied using invariant solutions \citep{chandler2013invariant,parker2022variational,page2024exact,page2024recurrent,redfern2024dynamically,cleary2025dynamical}. Previous searches explicitly targeted short periods \citep{page2024recurrent,redfern2024dynamically} but to our knowledge only three non-steady relative periodic orbits with $T<2$ have previously been found \citep{page2024exact,cleary2025dynamical,edinburgh2026orbits}. Comparisons also depend on the recurrence symmetry: a trajectory reported with a longer period may have a shorter recurrence modulo a discrete symmetry. Here we seek the shortest recurrence times across all symmetry classes described in \S\ref{sec:pos}. This allows us to ask whether the apparent scarcity of very short cycles reflects the dynamics or the limitations of previous searches.

Despite substantial progress in methods for converging periodic orbits \citep{ashtari2026computing}, obtaining suitable initial guesses remains a central challenge. Newton-based methods offer rapid local convergence, while recent variational approaches substantially enlarge the region from which convergence to an exact solution is possible \citep{lan2004variational,parker2022variational,azimi2022constructing}. (Here, \textit{exact} refers to numerical solutions to the governing equations subject to residual tolerance and resolution checks, rather than an analytic proof.) This progress is increasingly shifting the computational bottleneck from convergence itself to the generation of sufficiently informative and diverse initial guesses. How these guesses are generated also influences which invariant solutions are ultimately discovered. `Recurrent flow analysis' extract candidates from close returns observed along turbulent trajectories and consequently sample periodic orbits through the recurrences present in a finite simulation \citep{lucas2015recurrent,redfern2024dynamically}. Alternative guessing strategies introduce different sampling biases \citep{page2020searching,beck2024machine}. Thus, even though only a small proportion of the large number of known solutions are very short, this does not establish that such cycles are intrinsically rare.

Generative machine learning offers a different approach to the problem of generating initial guesses. Generative models learn statistical distributions from examples and can produce new samples that resemble the training data. Rather than waiting for a turbulent simulation to produce sufficiently close recurrences, a generative model trained on turbulent trajectories can be used to directly generate candidate recurrent motions. Importantly, these generated trajectories need not themselves accurately satisfy the governing equations. They only need to provide sufficiently plausible guesses to fall within the enlarged convergence region accessible to robust variational methods. The generative model explores the space of plausible recurrent trajectories, while more traditional numerical methods refine promising candidates into exact solutions of the governing equations.

Previous machine-learning approaches have assisted different stages of invariant-solution discovery. Autoencoders have provided equilibrium and travelling-wave guesses, latent recurrence detection and decoded periodic-orbit guesses in Kolmogorov flow \citep{page2021revealing,page2024exact}. Other recent approaches construct and glue latent-space loops \citep{beck2024machine,beck2026identifying}, locate orbits using a learned reduced model and a Poincar\'e map \citep{RN40}, or combine a patch-based autoencoder, learned time evolution and latent multiple shooting \citep{liu2026relative}.

Here, we use a diffusion model to synthesize entire candidate recurrent trajectories directly. We train on nonperiodic DNS segments, without examples of converged periodic orbits or a Navier--Stokes residual in the training objective. Recurrence, including closure up to symmetry transformations, is imposed during generation without retraining. These synthetic loops are then passed to a robust parallel-in-time convergence method and refined into solutions of the Navier--Stokes equations.

Using this approach, we identify 111 previously unknown relative periodic orbits with periods \(T<3\), including 40 with \(T<2\), substantially expanding the known population of very short cycles in this system. Their geometrically simple loops indicate that they represent a sample from a very large population of elementary recurrent motions, each contributing a single symbol to any possible symbolic description built from these motions. The resulting large alphabet makes a systematic cycle expansion computationally prohibitive with current methods: even a truncation at symbolic length two entails a potentially vast collection of combinations of these elementary cycles.

The remainder of the paper is organized as follows. In \S\ref{sec:kolmogorov} and \S\ref{sec:pos}, we introduce two-dimensional Kolmogorov flow, its symmetries and relative periodic orbits; \S\ref{sec:generativemodels} gives background on diffusion models. In \S\ref{sec:methods}, we describe the training data, the equivariant diffusion model and its adaptation for generating recurrent trajectories, followed by the numerical refinement method in \S\ref{sec:convergence}. Results for the full state space and an invariant subspace are presented in \S\ref{sec:results}, followed by the discussion in \S\ref{sec:discussion}.

\section{Background}
\subsection{Two-dimensional Kolmogorov flow}
\label{sec:kolmogorov}
We study the two-dimensional, incompressible Navier-Stokes equations with a sinusoidal forcing,
\begin{align}
        \partial_t u + u \partial_x u + v \partial_y u &= - \partial_x p + \frac{1}{Re} \Delta u + \sin 4y,\label{eq:governing_PV1}\\
    \partial_t v + u \partial_x v + v \partial_y v &= - \partial_y p + \frac{1}{Re} \Delta v,\label{eq:governing_PV2}\\
    \partial_x u + \partial_y v &= 0,\label{eq:governing_incompressibility}
\end{align}
solved on a doubly-periodic domain $x\in[0,2\pi)$, $y\in[0,2\pi)$. This is one form of Kolmogorov flow, which has been used numerous times as a model for turbulence in a computationally amenable configuration \citep{platt1991investigation, farazmand2017variational, lucas2022stabilization,de2023data, racca2023predicting}. Similar flows have also been studied experimentally \citep{suri2014velocity,suri2018unstable}. It is important to stress that, although these are indeed the incompressible Navier-Stokes equations, the resulting 2D spatio-temporal chaos is much simpler and quite different in behaviour from fully-developed 3D turbulence.

The equations given above are in the `primitive variables' formulation, describing the horizontal and vertical components of the fluid velocity as $u$ and $v$ and the pressure as $p$. This form of the problem is both intuitive and computationally efficient \citep{parker2022variational}. However, for an incompressible two-dimensional flow, these three variables can be reduced to the single variable of vorticity $\omega=\partial_x v - \partial_y u$, so that $u=-\partial_y\Delta^{-1}\omega$, $v=\partial_x\Delta^{-1}\omega$ and $p=-\Delta^{-1}(u_xu_x+v_yv_y+2u_yv_x)$ with a suitably defined inverse Laplacian $\Delta^{-1}$. 
By integrating \eqref{eq:governing_PV1} and \eqref{eq:governing_PV2} over the full periodic domain, we see that the spatially averaged horizontal and vertical velocities are conserved quantities of the system, which we shall always take to be zero. This provides the boundary conditions for the inverse Laplacian. We arrive at the alternative streamfunction-vorticity formulation for the system:
\begin{align}
    \partial_t \omega + \partial_y \psi \partial_x \omega - \partial_x \psi \partial_y \omega &= \frac{1}{Re}\Delta \omega -4\cos 4y,\label{eq:vorticityeq}\\
    \psi &= -\Delta^{-1} \omega.\label{eq:streamfunction}
\end{align}
Here, the flow at a given time is precisely defined by just one field, the vorticity (while also introducing the streamfunction $\psi$ -- a dummy variable -- for notational clarity).
This single variable form will be used throughout, in the direct solution of the evolution equations, and also to train the neural network described in section \ref{sec:diffusionmodel}. In a previous study \citep{parker2022variational}, we showed that the primitive variables formulation was more efficient for converging periodic orbits, but in the present work we employ a different method with an explicit Jacobian, and so to reduce memory consumption, we also work directly with the vorticity field in the convergence of periodic orbits, as discussed below.

Using the vorticity, the rate of energy dissipation and production are conveniently defined as, respectively,
\begin{equation}
    D = \frac{8}{Re^2\pi^2}\int_0^{2\pi}\int_0^{2\pi} \omega^2\,\mathrm{d}x\mathrm{d}y
\end{equation}
and
\begin{equation}
    P = -\frac{2}{Re\pi^2}\int_0^{2\pi}\int_0^{2\pi} \omega\cos{4y}\,\mathrm{d}x\mathrm{d}y,
\end{equation}
where, following previous authors, these quantities have been normalized so that $D=P=1$ for the laminar solution $\omega = -\frac{Re}{4}\cos4y$. The energy dissipation and production must balance over time, so that, for any exact solution to the equations, the time average of $D$ equals the time average of $P$.

\begin{figure}
    \centering
    \includegraphics[width=0.7\linewidth]{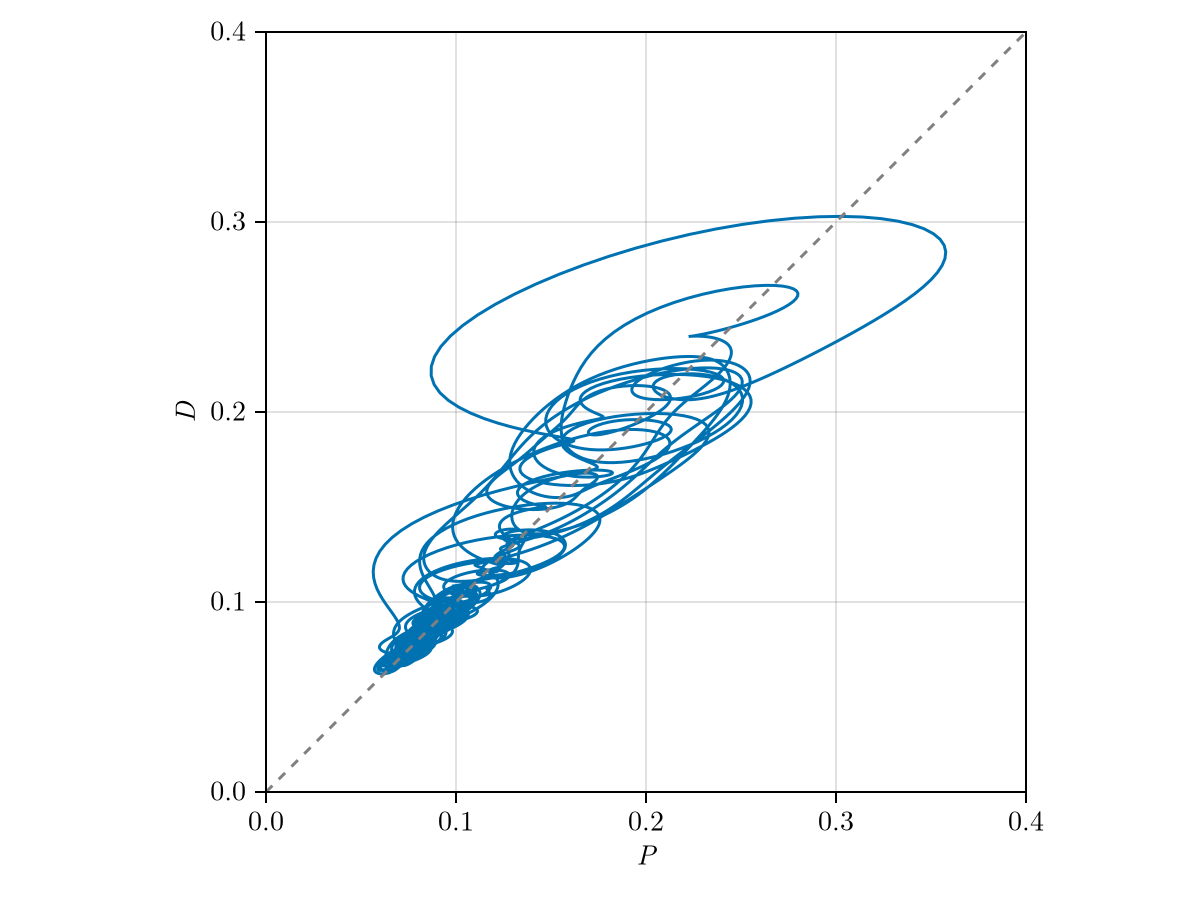}
    \caption{The rate of energy dissipation $D$ against production $P$ for a typical chaotic trajectory of length $T=500$. The trajectory spends the overwhelming majority of the time at low dissipations, with $D\le0.15$, but occasional `bursts' to higher $D$ are visible.}
    \label{fig:dpgeneric}
\end{figure}

Throughout, we take $Re=40$. This follows previous studies \citep{chandler2013invariant, parker2022variational, page2024recurrent, redfern2024dynamically}, and is modestly above the onset of spatiotemporal chaos at $Re\approx 31.5$. Previous authors have also studied higher values, up to $Re=1000$ \citep{cleary2025dynamical}. However, since the dynamics have not yet been fully captured at $Re=40$, it seems unnecessary to expend the significant computational resources required to accurately probe higher values until we have a strategy which works at $Re=40$. Furthermore, some evidence suggests that the dynamics is actually more complicated at $Re=40$ than at higher values \citep{cleary2025dynamical}. One notable feature of the dynamics in this regime is that states spend the majority of their time at low values of dissipation, but occasionally burst to much higher values, as visible in \cref{fig:dpgeneric}.

To solve \cref{eq:vorticityeq,eq:streamfunction}, we follow \citet{chandler2013invariant} in employing Heun's method with the Crank-Nicolson discretisation of the linear dissipative term. We use variously $N=64$, $N=96$ or $N=128$ Fourier modes in both the $x$ and $y$ directions, depending on the desired accuracy.

A symmetry of the system \eqref{eq:vorticityeq}-\eqref{eq:streamfunction} is a transformation of vorticity fields $\omega\mapsto\omega'$ such that given any solution $\omega(x,y,t)$ to the equations, $\omega'(x,y,t)$ is also a solution. 
With this particular choice of forcing, the system is invariant with respect to three distinct transformations, which together generate the symmetry group for the system. These are rotation through $\pi$,
\begin{equation}
    \mathcal{R}:\omega(x,y)\mapsto \omega(-x,-y),
\end{equation}
a shift in the $y$ direction by $\pi/4$, reflection in $x$ and a change of sign,
\begin{equation}
    \mathcal{S}:\omega(x,y)\mapsto -\omega(-x,y+\pi/4),
\end{equation}
and the continuous symmetry defined by a shift by an arbitrary distance $s$ in the $x$ direction (modulo the periodic boundary conditions)
\begin{equation}
    \mathcal{T}^s:\omega(x,y)\mapsto \omega(x+s,y).
\end{equation}
These have inverses $\mathcal{R}^{-1}=\mathcal{R}$, $\mathcal{S}^{-1}=\mathcal{S}^7$ and $\left(\mathcal{T}^s\right)^{-1}=\mathcal{T}^{-s}$. Note that these transformations do not commute in general. 
Any element in the symmetry group can be expressed uniquely in the form
$\mathcal{T}^s \mathcal{R}^a \mathcal{S}^m$ where $s\in [0,2\pi)$, $a\in\{0,1\}$ and $m\in \{0,\dots,7\}$.  More details can be found in \cref{sec:symmetries}.

Symmetries affect our problem in several ways: Firstly, given any generic solution, periodic or otherwise, we can find other, equivalent solutions by applying the symmetry operations. We can also find solutions which are periodic modulo the symmetry, as discussed in the next section. Finally, a solution can itself be symmetric.
If a vorticity field is invariant under a symmetry transformation, then the trajectory evolving from this field maintains that invariance for all time. Considering all possible symmetries in the group gives rise to an infinite number of invariant subspaces -- subsets of the full state space that trajectories can never enter or leave. We will mainly consider the full state space, but \cref{sec:Rresults} examines the solutions that are invariant under $\mathcal{R}$, i.e.\ vorticity fields invariant under rotation by $\pi$. This is analogous to studying antisymmetric solutions in another common model system for POs, the Kuramoto-Sivashinsky equation \citep{cvitanovic2010state,azimi2022constructing}. Authors often concentrate on antisymmetric solutions to reduce the complexity of the problem, even though these solutions are not a priori relevant to the chaotic attractor of the unconstrained system.

\subsection{Periodic orbits}
\label{sec:pos}
A \textit{periodic orbit} (PO) of the system \eqref{eq:vorticityeq}-\eqref{eq:streamfunction} is a non-stationary trajectory $\omega(x,y,t)$ such that $\omega(x,y,t+T)=\omega(x,y,t)$ for all $x$, $y$ and $t$, for some $T>0$. If the vorticity field is constant for all $t$ then this is instead an \textit{equilibrium}, and given the existence of the continuous symmetry, we also discount trajectories for which
\begin{equation}
    \omega(x,y,t) = \omega(x-ct,y,0)
\end{equation}
for some $c\ne0$ and all $t$, which are called \textit{travelling waves} (TWs) with respect to the shift symmetry $\mathcal{T}^s$. Given the existence of the continuous symmetry in this system, true periodic orbits are expected to be vanishingly rare \citep{cvitanovic2010state}. Instead, we expect \textit{relative periodic orbits} (RPOs), which are trajectories such that
\begin{equation}
    \omega(x,y,t+T) = \mathcal{G}\omega(x,y,t)
\end{equation}
for some symmetry $\mathcal{G}$. We say that this trajectory is an RPO relative to $\mathcal{G}$ with period $T$. If $\mathcal{G}$ is a discrete symmetry then this trajectory is also a full periodic orbit with period given by some integer multiple of $T$ (hence some authors refer to these as `pre-periodic' orbits), but we always focus on the shortest possible $T$, and do not draw a distinction between pre-periodic and relative periodic orbits.

Though the fundamental discrete symmetries of the system form a group of order 16, it is in fact sufficient to consider only 7 different discrete symmetry classes of RPOs. This is due to the fact that an RPO relative to one symmetry can be mapped onto another by applying some other symmetry transformations. For example, if there is some initial condition $\omega(x,y,0)$ such that after a time $T$ it satisfies 
\begin{equation}
\omega(x,y,T) = \mathcal{S}^6 \omega(x,y,0),
\end{equation}
then by letting $\omega'(x,y,0) = \mathcal{R}\omega(x,y,0)$, we have
\begin{equation}
\omega'(x,y,T) = \mathcal{S}^2 \omega'(x,y,0),
\end{equation}
so we do not need to consider $\mathcal{S}^6$ as well as $\mathcal{S}^2$.
We will only search for RPOs relative to $\mathcal{T}^s \mathcal{R}^a \mathcal{S}^m$ 
with
\[
\begin{aligned}
(a,m)&\in\{(0,0),(0,1),(0,2),(0,3),(0,4),(1,0),(1,1)\},\\[0.5ex]
s&\in
\begin{cases}
[0,2\pi), & a+m \text{ even},\\
\{0\},    & a+m \text{ odd}.
\end{cases}
\end{aligned}
\]
A full justification is given in \cref{sec:symmetries}.

It is important to emphasize the difference between an RPO relative to some symmetry and a PO in a symmetric subspace. The two can be combined -- there are RPOs in symmetric subspace.
In general, the existence of a continuous symmetry means that RPOs are crucial for understanding the dynamics \citep{cvitanovic2010state}, but when restricted to a symmetric subspace, this is no longer the case \citep{azimi2022constructing}.

\subsection{Diffusion models}
\label{sec:generativemodels}

Diffusion models are a family of generative machine learning approaches originally conceived to generate images. Unlike the earlier methods of generative adversarial networks \citep{goodfellow2014generative} and variational autoencoders \citep{kingma2013auto}, one single neural network does not generate a full image. Instead, a convolutional neural network is trained to predict the pattern of noise which has been added to a set of noise-free training images.
Conceptually, a diffusion model consists of two parts: a forward process, where random noise is incrementally added to the data following some predefined statistical process, and a reverse process, where a neural network, trained on a large dataset, predicts the noise which has been added, and which can then be subtracted to reproduce the original clean image. 
Then, starting instead with pure noise as input, the reverse process can be used to generate entirely synthetic images which resemble those in the training set.
Introduced by \citet{sohl2015deep}, diffusion models were initially overlooked due to computational inefficiencies but have since been revitalized through major innovations in training and sampling \citep{ho2020denoising}.

Different variants of diffusion models have been developed to address challenges such as computational expense and sampling efficiency.
Denoising Diffusion Probabilistic Models (DDPMs), popularised by \citet{ho2020denoising}, refine the reverse process stochastically, producing high-quality samples at the cost of requiring thousands of individual sampling steps. Their robust training dynamics and strong performance have made them a benchmark for diffusion-based generative models. Denoising Diffusion Implicit Models (DDIMs)  \citep{song2020denoising} improve upon DDPMs by reformulating the reverse process to allow deterministic sampling. This not only reduces the number of steps required for data generation but also introduces more control over the sampling process. We employ a DDIM in this work.

In the forward process, normally distributed noise $\epsilon \sim \mathcal{N}(0,1)$ is added to the original data $z^0$ via a fixed decreasing $\alpha_\tau\in(0,1]$ depending on the `timestep' $\tau$,
\begin{equation}
    z^{\tau} = \sqrt{\alpha_\tau} z^0 + \sqrt{1-\alpha_\tau}\epsilon,
\end{equation}
up to some final time $\tau=N_\tau$ for which $\alpha_\tau\approx0$, giving almost pure noise. This forward process is identical to that of DDPMs.
At each timestep of the reverse process, the model predicts the denoised version of the data, $\hat{z}^0$, from the noisy intermediate state $\hat{z}^\tau$ as
\begin{equation}
\hat{z}^0 = \frac{\hat{z}^\tau - \sqrt{1 - \alpha_\tau} \hat\epsilon(\hat{z}^\tau, \tau)}{\sqrt{\alpha_\tau}},
\end{equation}
where $\hat\epsilon$ is a neural network that estimates the added noise given the noisy data.
Using this estimate, DDIMs predict the next state in the denoising process,
\begin{equation}
\hat{z}^{\tau-\delta\tau} = \sqrt{\alpha_{\tau-\delta\tau}} \hat{z}^0 + \sqrt{1 - \alpha_{\tau-\delta\tau}} \hat\epsilon(\hat{z}^\tau, \tau),
\end{equation}
where $\delta\tau$ is the number of timesteps between each denoising iteration.

\section{Methods}
\label{sec:methods}
\subsection{Data generation}
\label{sec:data}
A long time-series of the chaotic attractor was computed at $Re=40$ using a spatial resolution of $N\times N=128\times128$, and a timestep of $\delta t=0.005$, using the numerical scheme mentioned in \cref{sec:kolmogorov}. For training the diffusion model, these data were downsampled to $N\times N =64\times64$ and $\delta t=0.02$, since the very fine scales required for accurate solution of the Navier-Stokes equations are presumed to be unimportant for the large-scale patterns we are trying to imitate.

The training data was comprised of 500 blocks of length $T=10.24$, corresponding to $M=512$ slices, which gives approximately 8GB of training data in uncompressed form.

Initial attempts to find periodic orbits resulted in a very large number of steady-state equilibrium solutions. This is likely due to the fact that, for most of the time, the Kolmogorov flow remains very close to the low-dissipation E1 solution \citep{chandler2013invariant}. To mitigate this, we instead trained our diffusion model only on high-dissipation events. Specifically, the time-series -- after an initial spin-up period of $T=1000$ was discarded -- was split into blocks of length $T=10.24$. Note that this is significantly longer than the individual periodic orbits we are targeting, allowing time for the dynamics to explore multiple regimes in a single block.
A block was only included in our training data if, somewhere within this block, the instantaneous dissipation was measured to be $D>0.15$. This cut-off value of $D=0.15$ was used by \citet{page2021revealing} to detect high-dissipation `bursting' events, and by this definition, all of our training data therefore contains a burst.

\subsection{Equivariant DDIM}
\label{sec:diffusionmodel}
Aside from the specific neural network architecture detailed below for our application, we used a standard DDIM \citep{song2020denoising}. 
The diffusion schedule has $N_\tau=1000$ noise levels. Generation uses 50 evenly spaced levels, corresponding to zero-based implementation indices $980,960,\ldots,0$, and performs 49 deterministic updates between them. Following \citet{ho2020denoising}, the variance schedule $\beta_\tau$ increases linearly from $\beta_1=10^{-4}$ to $\beta_{1000} = 0.02$, and then $\alpha_\tau$ is calculated as
\begin{equation}
    \alpha_\tau = \prod_{s=1}^{\tau}(1-\beta_s).
\end{equation}

In order to be able to generate time-series with different numbers of timesteps, and potentially different spatial sizes, we use a network whose field-processing path is fully convolutional. The dense layers act only on the diffusion-timestep embedding and do not fix the field resolution. One option would be a simple stack of convolutions which all preserve the size of the input. However, in order to more efficiently model large-scale features, we use a U-Net architecture \citep{ronneberger2015u}, which downsamples and then subsequently upsamples the resolution in both of the two spatial dimensions and in time. The pooling stages reduce the spatial resolution by a factor of 8 and the temporal resolution by a factor of 4, requiring $N$ to be divisible by 8 and $M$ by 4. To represent the $\pi/4$ shift in $\mathcal{S}$ exactly on every spatial grid additionally requires $N$ to be divisible by 64; we use $N=64$ for the diffusion model.

The vorticity fields are rescaled by a factor of 10 for the diffusion model, which has the effect of keeping vorticity values in the range $[-1,1]$. Let a noisy field in these scaled units be denoted $z_{ijk}$, where $1\leq i \leq N$, $1\leq j \leq N$ and $1 \leq k \leq M$.
The neural network is a function $\hat\epsilon:\mathbb{R}^{N\times N\times M}\times \mathbb{N}\to \mathbb{R}^{N\times N\times M}$. At each selected diffusion level, the DDIM forms a clean-field estimate 
\begin{equation}
\hat{z}^0 = \frac{\hat z^\tau - \sqrt{1 - \alpha_\tau} \hat\epsilon(\hat z^\tau, \tau)}{\sqrt{\alpha_\tau}},
\end{equation}
which is clipped to $[-1,1]$ before the deterministic update to the next selected level. The implementation returns the state at its last selected noise level (zero-based index 0, or $\tau=1$ in the convention above), without an additional noise-free prediction. Denoting this returned field by $\hat z_{\mathrm{out}}$, the initial guess for numerical refinement is
\begin{equation*}
    \omega\left((2i-1)\pi/N,(2j-1)\pi/N,k\delta_t\right) = 10(\hat z_{\mathrm{out}})_{i,j,k}.
\end{equation*}
No explicit spatial-mean subtraction is applied to the network output.

A fully-convolutional network can accommodate different input sizes, subject to the divisibility constraints above. Its translation equivariance is restricted by the pooling operations.
The design of the network is strongly constrained by the fact that we wish to be able to generate periodic orbits, and relative periodic orbits with various symmetries: suppose hypothetically that the input to the network were an infinitely long sequence of $N\times N$ slices, so that $\hat\epsilon : \mathbb{R}^{N\times N \times \mathbb{Z}}\times \mathbb{N}\to \mathbb{R}^{N\times N \times \mathbb{Z}}$. Since the network downsamples the inputs by a factor of 4, it cannot be the case that shifting the input by 1 in the time direction results in the same output but perfectly shifted by 1. However, by careful choice of the convolution, downsampling and upsampling operations, we can ensure that shifting the inputs by 4 in time gives an identical output, but shifted by 4. That is to say, letting $z_{i,j,k}' = z_{i,j,k+4}$, we have
\begin{equation}
    \left(\hat\epsilon\left(z',\tau\right)\right)_{i,j,k} = \left(\hat\epsilon\left(z,\tau\right)\right)_{i,j,k+4}.
\end{equation}
In order for this to hold, all convolutions have a stride of 1. The usual max-pooling for downsampling would break the symmetry $\mathcal{S}$ we wish to preserve, so instead we use average pooling with sizes $2\times2\times1$ and $2\times2\times2$, and corresponding upsampling with a custom trilinear interpolation. With these choices, suppose that the input were periodic with period $4K$ for some $K$, i.e. $z_{i,j,k+4K} = z_{i,j,k}$. Then
\begin{equation}
    \left(\hat\epsilon\left(z,\tau\right)\right)_{i,j,k+4K} = \left(\hat\epsilon\left(z,\tau\right)\right)_{i,j,k},
\end{equation}
for every $\tau$
and therefore the generated vorticity field satisfies
\begin{equation}
    \omega(x,y,t+4K\delta_t) = \omega(x,y,t),
\end{equation}
a synthetic periodic orbit with period $4K\delta_t$. 
Furthermore, we wish to preserve the symmetries of the system described in section \ref{sec:kolmogorov} that are represented exactly on the network's nested grids. For each such element $\mathcal{G}$, the function $\hat\epsilon$ should be equivariant with respect to its action on inputs $z$
\begin{equation}
    \hat\epsilon(\mathcal{G}z,\tau) = \mathcal{G} \hat\epsilon(z,\tau),
\end{equation}
for all $z$ and $\tau$.

One important consideration for this to hold is how we handle the input $\tau$, which is the current `timestep' of the diffusion process. (This is unrelated to the timestep of the spatiotemporal field $z$, which encodes time in its third dimension.) Typically in DDPMs and DDIMs this is done with some form of positional encoding, which encodes the integer $\tau$ as a series of sines and cosines, which are then input into a small multilayer perceptron (MLP) which can learn a suitable representation of this for the network \citep{vaswani2017attention, ho2020denoising}. We follow this procedure, multiplying the embedded timestep by the scaled forcing profile $f_{i,j,k} = \sqrt{2}\cos\left(\frac{4(2j-1)\pi}{N}\right)$ evaluated on each block's grid and concatenating the resulting channels onto the inputs to that block. 
Using the forcing field in this way has the added benefit of removing the additional $y$-translation equivariances that the convolution and pooling operations would otherwise admit, but which are not symmetries of the forced equations.

For horizontal shifts, the streamwise translation acts as
\begin{equation*}(\mathcal{T}^s z)_{i,j,k} = z_{i+\frac{sN}{2\pi},j,k}.\end{equation*}
Here, indices outside $1\leq i\leq N$ are understood periodically. Individual stride-one convolutions with periodic padding commute with these shifts. For the full U-Net, the three spatial pooling stages restrict the exact translation equivariance to shifts by multiples of eight input cells: $s\in(16\pi/N)\mathbb{Z}$, or multiples of $\pi/4$ at $N=64$. Nevertheless, the use of average-pooling means that horizontal shifts by other amounts are still an approximate equivariance.
We also use periodic padding in the $y$ direction, but shifts by arbitrary amounts in this direction are not equivariant, since the constant forcing field $f$ is not itself invariant under arbitrary $y$ shifts.

The rotation symmetry $\mathcal{R}$ is also straightforward to handle: all kernels of the convolution layers must themselves be invariant under a rotation through $\pi$. 
The shift-and-reflect symmetry $\mathcal{S}$ is somewhat more complicated to incorporate into the neural network architecture. For $\mathcal{S}^m$ with $m$ even, it reduces to a translation in $y$ by $mN/8$ grid cells. At $N=64$, these even-$m$ shifts are compatible with every pooling level, and the forcing field satisfies $\mathcal{S}^2 f = f$. However, when $m$ is odd, the symmetry introduces a change of sign to the output. One option would be for all activation layers $\sigma$ in the neural network to satisfy $\sigma(-z_{ijk})=-\sigma(z_{ijk})$, and this, combined with the fact that all other layers (convolutions, average pooling and trilinear upsampling) are purely linear, would mean that $\hat\epsilon(-z)=-\hat\epsilon(z)$. But pure change-of-sign is not a symmetry of \cref{eq:vorticityeq} -- it is not the case that given a solution $\omega(x,y,t)$, another is given by $-\omega(x,y,t)$. Therefore, such a choice would put an additional restriction on the neural network which is unnatural and would lead to poor performance and generalisation. Instead, we must take into account the reflection in $x$.

To proceed, let us examine the individual terms of (\ref{eq:vorticityeq}). The time derivative and Laplacian terms are purely linear in $\omega$, and thus equivariant under a change of sign. Additionally, these terms are equivariant under reflection in $x$, as the former has no spatial derivatives and the latter has a second derivative in $x$. 
The nonlinear terms $\partial_y \psi \partial_x \omega$ and $\partial_x \psi \partial_y \omega$ are completely different: they are \textit{invariant} under a change of sign of $\omega$ (since $\psi=-\Delta^{-1}\omega$ changes sign with $\omega$) and change sign under reflection in $x$, since each term has one derivative in $x$. This inspires handling these two modes of symmetry in parallel in our neural network: either we have (a) a convolution which is symmetric in $x$ and an activation which is an odd function (i.e. equivariant under a change of sign) or we have (b) an even function followed by a convolution which is antisymmetric in $x$ and then an odd activation function. 
Type (a) operations are equivariant under reflection in $x$ and under sign reversal. Type (b) operations are invariant under input sign reversal, while reflecting the input produces the negative of the reflected output.
Both types are therefore equivariant under the combined operation of reflection in $x$ and sign reversal. Combining them does not require equivariance under either operation separately. Together with the compatible $y$-shift and the forcing input, which satisfies $\mathcal{S}f=f$, this provides the symmetry structure needed for $\mathcal{S}$. Preserving this equivariance through the full block also requires zero additive biases and normalization that commutes with the symmetry.

\begin{figure}
    \centering
    \begin{tikzpicture}[line join=round]
    \begin{scope}[xshift=0.0cm]
        \foreach \depth in {4,3,2,1,0} {
            \begin{scope}[shift={({0.25*\depth},{0.22*\depth})}]
                \filldraw[fill=white,draw=black!65,line width=0.5pt] (0,0) rectangle (3,3);
                \draw[step=0.6cm,draw=black!25,line width=0.3pt] (0,0) grid (3,3);
            \end{scope}
        }
        \node[font=\small] at (0.3,2.7) {$a_1$};
        \node[font=\small] at (0.9,2.7) {$a_2$};
        \node[font=\small] at (1.5,2.7) {$a_3$};
        \node[font=\small] at (2.1,2.7) {$a_2$};
        \node[font=\small] at (2.7,2.7) {$a_1$};
        \node[font=\small] at (0.3,2.1) {$a_4$};
        \node[font=\small] at (0.9,2.1) {$a_5$};
        \node[font=\small] at (1.5,2.1) {$a_6$};
        \node[font=\small] at (2.1,2.1) {$a_5$};
        \node[font=\small] at (2.7,2.1) {$a_4$};
        \node[font=\small] at (0.3,1.5) {$a_7$};
        \node[font=\small] at (0.9,1.5) {$a_8$};
        \node[font=\small] at (1.5,1.5) {$a_9$};
        \node[font=\small] at (2.1,1.5) {$a_8$};
        \node[font=\small] at (2.7,1.5) {$a_7$};
        \node[font=\small] at (0.3,0.9) {$a_4$};
        \node[font=\small] at (0.9,0.9) {$a_5$};
        \node[font=\small] at (1.5,0.9) {$a_6$};
        \node[font=\small] at (2.1,0.9) {$a_5$};
        \node[font=\small] at (2.7,0.9) {$a_4$};
        \node[font=\small] at (0.3,0.3) {$a_1$};
        \node[font=\small] at (0.9,0.3) {$a_2$};
        \node[font=\small] at (1.5,0.3) {$a_3$};
        \node[font=\small] at (2.1,0.3) {$a_2$};
        \node[font=\small] at (2.7,0.3) {$a_1$};
        \draw[->,draw=black!65] (0,3.3) -- (1,4.18) node[midway,above,sloped,font=\small] {time};
        \node at (1.5,-0.4) {(a) Symmetric};
    \end{scope}
    \begin{scope}[xshift=5.2cm]
        \foreach \depth in {4,3,2,1,0} {
            \begin{scope}[shift={({0.25*\depth},{0.22*\depth})}]
                \filldraw[fill=white,draw=black!65,line width=0.5pt] (0,0) rectangle (3,3);
                \draw[step=0.6cm,draw=black!25,line width=0.3pt] (0,0) grid (3,3);
            \end{scope}
        }
        \node[font=\small] at (0.3,2.7) {$b_1$};
        \node[font=\small] at (0.9,2.7) {$b_2$};
        \node[font=\small] at (1.5,2.7) {$0$};
        \node[font=\small] at (2.1,2.7) {$-b_2$};
        \node[font=\small] at (2.7,2.7) {$-b_1$};
        \node[font=\small] at (0.3,2.1) {$b_3$};
        \node[font=\small] at (0.9,2.1) {$b_4$};
        \node[font=\small] at (1.5,2.1) {$0$};
        \node[font=\small] at (2.1,2.1) {$-b_4$};
        \node[font=\small] at (2.7,2.1) {$-b_3$};
        \node[font=\small] at (0.3,1.5) {$0$};
        \node[font=\small] at (0.9,1.5) {$0$};
        \node[font=\small] at (1.5,1.5) {$0$};
        \node[font=\small] at (2.1,1.5) {$0$};
        \node[font=\small] at (2.7,1.5) {$0$};
        \node[font=\small] at (0.3,0.9) {$-b_3$};
        \node[font=\small] at (0.9,0.9) {$-b_4$};
        \node[font=\small] at (1.5,0.9) {$0$};
        \node[font=\small] at (2.1,0.9) {$b_4$};
        \node[font=\small] at (2.7,0.9) {$b_3$};
        \node[font=\small] at (0.3,0.3) {$-b_1$};
        \node[font=\small] at (0.9,0.3) {$-b_2$};
        \node[font=\small] at (1.5,0.3) {$0$};
        \node[font=\small] at (2.1,0.3) {$b_2$};
        \node[font=\small] at (2.7,0.3) {$b_1$};
        \draw[->,draw=black!65] (0,3.3) -- (1,4.18) node[midway,above,sloped,font=\small] {time};
        \node at (1.5,-0.4) {(b) Antisymmetric};
    \end{scope}
    \end{tikzpicture}
    \caption{Spatial symmetry of the convolution kernels of type (a) and (b), shown as stacks of five $5\times5$ matrices along the time direction. The front slices display the spatial coefficient pattern, with respectively 9 and 4 independent coefficients per slice. The other four slices have the same spatial symmetry but independent coefficients, giving 45 and 20 parameters per input--output channel pair. No symmetry is imposed in time or between channel pairs. The full kernel arrays have size $5\times5\times5\times d_i\times d_o$.}
    \label{fig:kernels}
\end{figure}

Convolutions of type (a) have kernels which are symmetric in the first dimension and also symmetric under rotation by $\pi$, as required for equivariance under $\mathcal{R}$. For a $5\times5$ filter this leaves 9 unique elements. Convolutions of type (b) have kernels anti-symmetric in the first dimension but still symmetric under rotation by $\pi$, which leaves just four unique element for $5\times5$ convolutions. This is depicted graphically in figure \ref{fig:kernels}. 
In practice, we also perform a convolution in time, with no symmetry restrictions in that direction, so for $5\times5\times5$ convolutions, we have 45 parameters for type (a) and 20 for type (b). Further to this, the convolutions are performed across a variable number of input channels and output channels, so the number of parameters rapidly becomes very large, though still much smaller than if these symmetry constraints were not imposed on the convolutions.

Activations functions after each layer contribute nonlinearity to an otherwise entirely linear neural network. Many architectures use ReLU activation functions or modifications thereof, which are computationally efficient and do not suffer from saturation. However, this family of functions are neither even nor odd, so cannot be used if we wish to preserve symmetries. The widely used hyperbolic tangent activation is an odd function, but suffers from vanishing gradients. For this reason we use an activation function $\mathrm{tanhshrink}(x)= x-\tanh{x}$  \citep{sipper2021neural} after each symmetric or antisymmetric convolution within a residual block. The $1\times1\times1$ residual projections and the final output convolution have no activation. For the even function required before the convolution in type (b) operations, we use a function $\mathrm{softabs}(x)= \frac{|x|^3}{0.1+x^2}$, which approximates the usual absolute value with an everywhere-differentiable function. 

Putting these together, the neural network is assembled from convolutional blocks containing parallel type (a) and (b) convolutions with their respective activations, concatenation, followed by a second pair of convolutions and concatenation. Performance was observed to increase with the inclusion of a residual connection past the convolutions which is then added to the output of the block, and also the use of group normalisation after each of the concatenations. The full convolutional block is depicted in \cref{fig:conv}. Group normalisation uses eight groups, a trainable scale and no additive offset. When the input and output channel counts agree, the residual connection is the identity; otherwise a $1\times1\times1$ projection matches the channel counts.

The diffusion-timestep embedding comprises 16 sine and 16 cosine components with logarithmically spaced frequencies from $1$ to $10^{-4}$. Two dense layers of width 32, with a swish activation after the first and no activation after the second, produce 32 coefficients. At the start of each convolutional block, these coefficients multiply samples of the forcing profile $\sqrt{2}\cos(4y)$ on that block's grid. The resulting 32 channels are constant in $x$ and time and are concatenated with the incoming field channels.

An implementation oversight left trainable biases in the six $1\times1\times1$ residual projections which skip the type (a) and (b) convolutions, making equivariance of the full neural network under sign-changing symmetries approximate rather than exact. A posteriori checks with unit-Gaussian inputs of size $64\times64\times64$ at zero-based diffusion index 500 gave an RMS difference between $\hat\epsilon(\mathcal{S}z,\tau)$ and $\mathcal{S}\hat\epsilon(z,\tau)$ of approximately $1.5\%$ and $2.7\%$ of the output RMS for periodic and $\mathcal{S}$-twisted temporal padding, respectively. Setting only these biases to zero reduced the relative difference to $6.3\times10^{-5}$ and $1.2\times10^{-5}$, respectively, and changed the network outputs by approximately $1.6\%$ in RMS. These bias terms do not break $\mathcal{R}$ or the sign-preserving symmetries. Their effect on orbit-discovery rates has not been measured. The reported orbits are nevertheless independently refined and checked against the governing equations and their recurrence conditions, so their validity does not rely on exact network equivariance.

\begin{figure}
    \centering
\begin{tikzpicture}[
  font=\small,
  >={Stealth[length=1.4mm,width=1mm]},
  box/.style={draw, line width=.4pt, align=center, inner sep=3pt},
  op/.style={box,text width=2.9cm,minimum height=1.7cm},
  norm/.style={box,text width=3cm,minimum height=.75cm},
  flow/.style={->,line width=.4pt},
  wire/.style={line width=.4pt},
  cond/.style={flow,dash pattern=on 2pt off 2pt}
]

\node (in) at (4.3,0) {$64^3\times48$};
\node[box,text width=2.3cm] (time) at (7.7,0)
  {$\tau$-embedding\\[2pt]32 channels};
\node[box,text width=2.5cm,minimum height=.5cm] (cat) at (4.3,-1.15)
  {Concat: 80};
\draw[flow] (in.south) -- (cat.north);
\draw[cond] (time.south) |- (cat.east);

\node[op] (a1) at (2.3,-2.75)
  {Type (a) conv\\[2pt]$80\to8$\\[2pt]tanhshrink};
\node[op] (b1) at (6.3,-2.75)
  {softabs\\[2pt]Type (b) conv\\[2pt]$80\to8$\\[2pt]tanhshrink};
\coordinate (split1) at (4.3,-1.7);
\draw[wire] (cat.south) -- (split1);
\draw[flow] (split1) -| (a1.north);
\draw[flow] (split1) -| (b1.north);

\node[norm] (n1) at (4.3,-4.5)
  {Concat: 16\\[2pt]Group norm.};
\coordinate (join1) at (4.3,-3.85);
\draw[wire] (a1.south) |- (join1);
\draw[wire] (b1.south) |- (join1);
\draw[flow] (join1) -- (n1.north);

\node[op] (a2) at (2.3,-6.25)
  {Type (a) conv\\[2pt]$16\to8$\\[2pt]tanhshrink};
\node[op] (b2) at (6.3,-6.25)
  {softabs\\[2pt]Type (b) conv\\[2pt]$16\to8$\\[2pt]tanhshrink};
\coordinate (split2) at (4.3,-5.2);
\draw[wire] (n1.south) -- (split2);
\draw[flow] (split2) -| (a2.north);
\draw[flow] (split2) -| (b2.north);

\node[norm] (n2) at (4.3,-8)
  {Concat: 16\\[2pt]Group norm.};
\coordinate (join2) at (4.3,-7.35);
\draw[wire] (a2.south) |- (join2);
\draw[wire] (b2.south) |- (join2);
\draw[flow] (join2) -- (n2.north);

\node[box,text width=1.8cm,minimum height=.85cm] (res) at (-.7,-6.25)
  {$1^3$ conv\\[2pt]$48\to16$};
\draw[flow] (in.west) -| (res.north);
\node[draw,circle,line width=.4pt,inner sep=1.5pt,minimum size=.35cm]
  (add) at (4.3,-9.05) {$+$};
\draw[flow] (res.south) |- (add.west);
\draw[flow] (n2.south) -- (add.north);
\node (out) at (4.3,-9.7) {$64^3\times16$};
\draw[flow] (add.south) -- (out.north);
\end{tikzpicture}
    \caption{The convolutional block used throughout the neural network, shown for the final decoder block of \cref{fig:unet}. The symmetric and antisymmetric $5\times5\times5$ convolutions use kernels of type (a) and (b) in \cref{fig:kernels}, respectively. Spatial and temporal padding, omitted here, precede each of these convolutions. Each branch produces eight channels, followed by tanhshrink activation; the antisymmetric branch first applies softabs. Group normalisation uses eight groups, a trainable scale and no additive offset. The 32 forcing channels are concatenated only at the block entrance. The residual projection acts on the original 48 channels, has no activation and includes the additive bias discussed in the text. Blocks with matching input and output channel counts instead use an identity residual connection.}
    \label{fig:conv}
\end{figure}

The full structure of the U-Net for $\hat\epsilon$ is depicted in \cref{fig:unet}, and the full source code, implemented with Keras/Tensorflow with several custom layers, is available at \url{https://github.com/jeremypparker/Parker_Schneider_Diffusion_Model}.

\begin{figure}
    \centering
\begin{tikzpicture}[
  font=\small,
  >={Stealth[length=1.4mm,width=1mm]},
  block/.style={draw, line width=.4pt, align=center,
    text width=2.7cm, minimum height=.85cm, inner sep=3pt},
  flow/.style={->, line width=.4pt},
  cond/.style={line width=.4pt, dash pattern=on 2pt off 2pt},
  op/.style={font=\footnotesize, inner sep=2pt}
]

\node[block,text width=1.8cm,minimum height=.5cm]
  (time) at (3.5,-7.05) {Embedding};
\node (tau) at (1.8,-7.05) {$\tau$};
\draw[flow] (tau) -- (time);

\coordinate (bus) at (3.5,-6.05);
\draw[cond] (time.north) -- (bus);
\draw[cond] (bus) -- (-1.85,-6.05) -- (-1.85,-.25);
\draw[cond] (bus) -- (8.85,-6.05) -- (8.85,-.25);

\node[align=center] (in) at (0,1.65)
  {$\hat z^\tau$\\[2pt]$64^3\times1$};
\node[align=center] (out) at (7,1.65)
  {$\hat\epsilon(\hat z^\tau,\tau)$\\[2pt]$64^3\times1$};
\node[block,minimum height=.4cm] (last) at (7,.65) {$5^3$ conv};

\node[block] (e1) at (0,-.25) {Conv block\\[2pt]$64^3\times16$};
\node[block] (e2) at (0,-1.9) {Conv block\\[2pt]$32^2\times64\times32$};
\node[block] (e3) at (0,-3.55) {Conv block\\[2pt]$16^2\times32\times64$};
\node[block] (d1) at (7,-.25) {Concat + conv\\[2pt]$64^3\times16$};
\node[block] (d2) at (7,-1.9) {Concat + conv\\[2pt]$32^2\times64\times32$};
\node[block] (d3) at (7,-3.55) {Concat + conv\\[2pt]$16^2\times32\times64$};
\node[block,text width=3.4cm] (b) at (3.5,-5.15)
  {3 conv blocks\\[2pt]$8^2\times16\times64$};

\draw[flow] (in.south) -- (e1.north);
\draw[flow] (e1.south) -- node[op,right=4pt] {Pool $(2,2,1)$} (e2.north);
\draw[flow] (e2.south) -- node[op,right=4pt] {Pool $(2,2,2)$} (e3.north);
\draw[flow] (e3.south) -- node[op,right=4pt,pos=.2] {Pool $(2,2,2)$}
  (0,-5.15) -- (b.west);
\draw[flow] (b.east) -- (7,-5.15) --
  node[op,left=4pt,pos=.8] {Upsample $(2,2,2)$} (d3.south);
\draw[flow] (d3.north) -- node[op,left=4pt] {Upsample $(2,2,2)$} (d2.south);
\draw[flow] (d2.north) -- node[op,left=4pt] {Upsample $(2,2,1)$} (d1.south);
\draw[flow] (d1.north) -- (last.south);
\draw[flow] (last.north) -- (out.south);

\foreach \i in {1,2,3} {
  \draw[flow] (e\i.east) -- (d\i.west);
  \draw[cond,->] (-1.85,0 |- e\i.west) -- (e\i.west);
  \draw[cond,->] (8.85,0 |- d\i.east) -- (d\i.east);
}
\draw[cond,->] (3.5,-6.05) -- (b.south);
\end{tikzpicture}
    \caption{The structure of the U-Net used by the diffusion model. Each conv block is as per \cref{fig:conv}. There are three convolutional blocks at the bottleneck. The final layer is a bias-free symmetric $5\times5\times5$ convolution with no activation. Concatenation is along the fourth dimension. The sizes given are those used during generation, with repeated dimensions abbreviated using powers; during training, all layers were $8$ times larger in the third dimension.}
    \label{fig:unet}
\end{figure}

\subsubsection{Padding}
To preserve the size before and after a discrete convolution, the inputs must be padded. With the $5\times5\times5$ kernels we use, this means that two elements must be augmented to the data on either side of the first three dimensions. In all cases we use periodic padding in both spatial dimensions to represent the periodic boundary conditions of the PDE. That is, the first two rows/columns of each slice are appended to the opposite side, and likewise with the final two. In the third, temporal dimension, we must be more careful. The training data (discussed in \cref{sec:data}) is not periodic in time, and therefore using periodic padding would result in a discontinuity. We use mirror padding in the third dimension during training. That is, the last two slices are swapped and appended to the end, and similarly the first two slices at the start. There is no physical motivation for this choice, but it was found to give better results than the more common zero padding.

During the generation phase of our method, we aim to create synthetic trajectories which are explicitly periodic in time, or periodic relative to a transformation in the symmetry group of the system. In the former case, of non-relative periodic orbits, these can be generated by simply using normal periodic padding in the third dimension as in the first two dimensions. We only do this in the $\mathcal{R}$-invariant subspace as described in \cref{sec:Rresults}, in all other cases we are concerned with non-trivial symmetry transformations, which requires a custom padding operation. In cases where the transformation $\mathcal{T}^s \mathcal{R}^a \mathcal{S}^m$ has $s=0$, consisting only of rotations, reflections and discrete vertical shifts, this can be done exactly: We pad the end of the third dimension with the first two slices, transformed by $\mathcal{T}^s \mathcal{R}^a \mathcal{S}^m$, and we pad the start with the final two slices transformed by the inverse operation. When $s\ne0$, we apply the same construction, using linear interpolation wherever the shift is not aligned with a layer's grid. This introduces small interpolation errors for arbitrary shifts. At $N=64$, shifts by integer multiples of $\pi/4$ are grid-aligned at every pooling level and belong to the translation subgroup preserved by the network.

\subsubsection{Training}
The network was trained for 2500 epochs on 500 spatiotemporal blocks, each containing 512 snapshots on a $64\times64$ spatial grid. Vorticity was divided by 10 before applying the forward diffusion process. At each parameter update, a diffusion level was sampled uniformly from the 1000 levels, and a fresh Gaussian noise field was generated. We used the Adam optimizer with fixed learning rate $5\times10^{-5}$, a batch size of one, and a mean-squared-error loss for noise prediction. The training blocks were reshuffled at each epoch. Training took approximately six days on a single NVIDIA Tesla V100 GPU with 32\,GB of memory.

\subsection{Converging POs}
\label{sec:convergence}
The strength of our method is the combination of a tool for generating plausible closed trajectories with a tool for converging such closed loops to solutions of the governing equations. Many other studies on finding periodic orbits in this system have used Newton-shooting, which takes as input an initial flow configuration and a guess of the period $T$, and integrates the equations up to time $T$ with the hope that the result is close to the initial condition. The initial condition and period are then refined using Newton's method until the loop closes. Our approach instead iteratively modifies a given trajectory, which is always a closed loop in state space, until it satisfies the governing equations at each point. Every timestep is simultaneously updated in this `parallel-in-time' algorithm, with no preference given for which slice is the initial condition. The period $T$ and, where relevant, the horizontal shift $s$, are also simultaneously corrected.

The aim of the present work is to demonstrate a new method for generating guesses for periodic orbits, rather than to develop a new method for their convergence, since a great deal of published literature deals with this latter step \citep{viswanath2001lindstedt,lan2004variational,azimi2022constructing}. In particular, in \citet{parker2022variational} we described a method for converging relative periodic orbits from guesses given as loops in state space for this system. Nevertheless, to fully exploit the symmetries of the system and to take into account lessons learnt from \citet{parker2023predicting}, it was convenient to create a new code to converge the guesses generated by our neural network. The method, implemented in Julia for execution on either a CPU or GPU, uses an explicit Jacobian, stored as a sparse matrix, to converge a full space-time field by a Levenberg-Marquardt-like algorithm. 
This method aims to combine the large radius of convergence of gradient-based methods \citep{parker2022variational,page2024recurrent} with the quadratic convergence of Newton's method \citep{viswanath2007recurrent,chandler2013invariant} near a solution.
The relative periodic orbit is discretized by high-order filtered finite differences in time, which allows the imposition of the desired symmetries through a derivative matrix and gives a sparser Jacobian than using a spectral method. For spatial discretisation, we use Fourier modes with 2/3 dealiasing for the nonlinear terms, in order to exploit the periodicity of the domain. The source code is available at \url{https://github.com/jeremypparker/Parker_Schneider_Diffusion_Model}.

One subtlety is that our diffusion model acts on grids of datapoints, whereas our convergence algorithm operates entirely on Fourier coefficients in the two spatial directions, except for a pseudo-spectral collocation step to handle the nonlinearity. This collocation uses gridpoints at $x=0,2\pi/N,\dots$, allowing the symmetry operations to be performed efficiently in Fourier space. The neural network, by contrast, expects datapoints at $x=\frac{1}{2}\times2\pi/N, \frac{3}{2}\times2\pi/N,\dots$, which means, for example, that $\mathcal{R}$ is simply a rotation of the whole image. This necessitates a half-pixel shift via linear interpolation when the synthetic orbits are used as initial guesses for the convergence algorithm, which results in a small loss of fidelity.

\section{Results}
\label{sec:results}
\subsection{Full state space}
\label{sec:fullspaceresults}
For each of the 7 unique classes of RPO, we generated 400 synthetic orbits using the diffusion model, with $(N,M)=(64,64)$. 
Since $\delta t=0.02$, $M=64$ corresponds to RPOs of length $T=1.28$, which is less than every previously known RPO for this system. This was chosen in order to force the method to find the very shortest solutions.
On a single NVIDIA Tesla V100 GPU with 32 GB of memory, each guess took around 8 seconds to generate.

\begin{figure}
    \centering
    \includegraphics[width=0.7\linewidth]{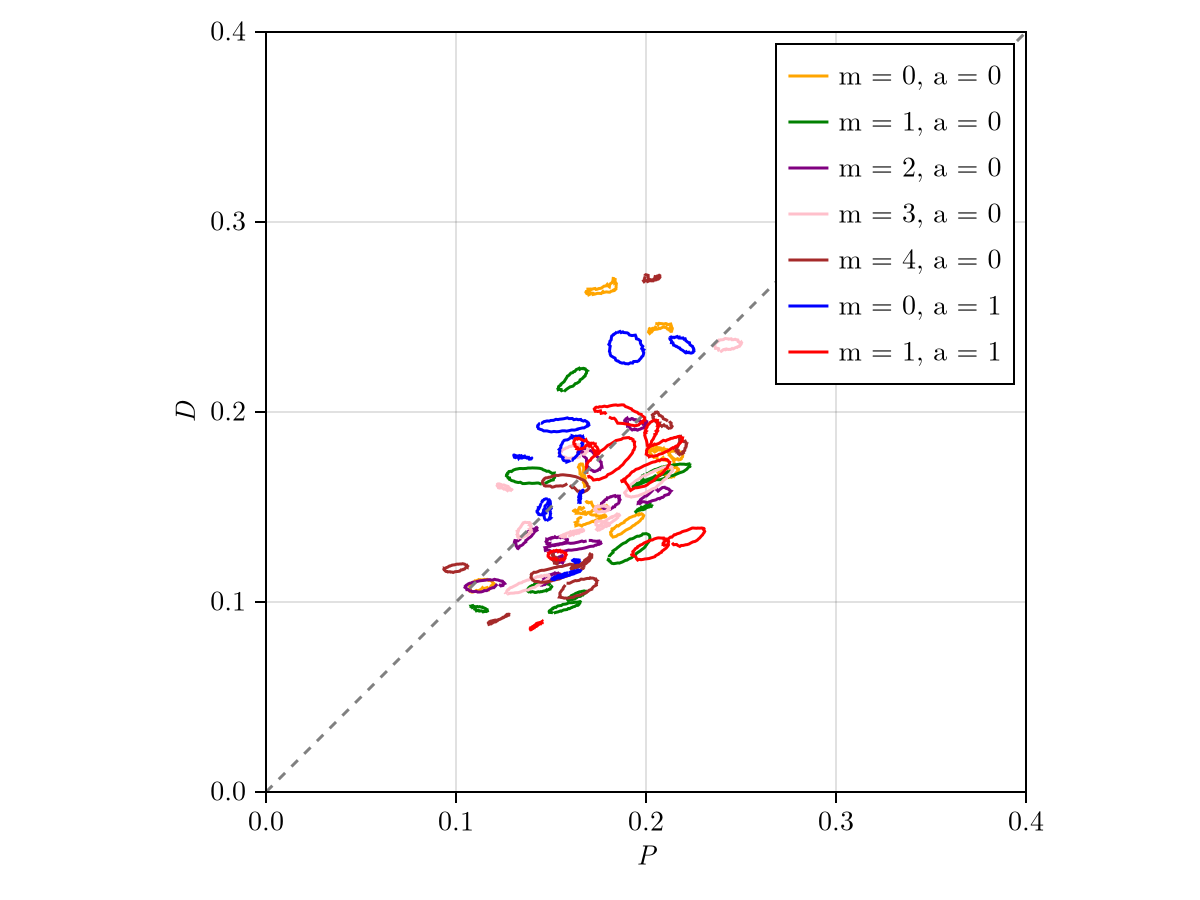}
    \caption{Energy dissipation $D$ versus production $P$ for synthetic orbits generated with $(N,M)=(64,64)$. Ten examples from each of the symmetries are shown. The examples were chosen at random: many of these converged to true solutions, many did not.}
    \label{fig:examples}
\end{figure}

Examples of the resulting synthetic orbits are plotted in \cref{fig:examples}.
These show only a very weak correlation, if any, between the average values of $D$ and $P$, which must be equal for any true solution. However, there is clear evidence that the network has learnt that dissipation varies much less than production on any given orbit, as visible in the short and wide oval shapes.
All our synthetic orbits have medium values of dissipation, and there is no clear pattern between the synthetic orbits of the different symmetries. Crucially, we have not been able to generate orbits of very low dissipation which have historically been the most common ones found by previous studies. This is partly due to the short periods we are targeting, and the fact that our model was trained only on high-dissipation bursting events, as discussed in \cref{sec:data}.

\begin{figure}
    \centering
    \includegraphics[width=0.7\linewidth]{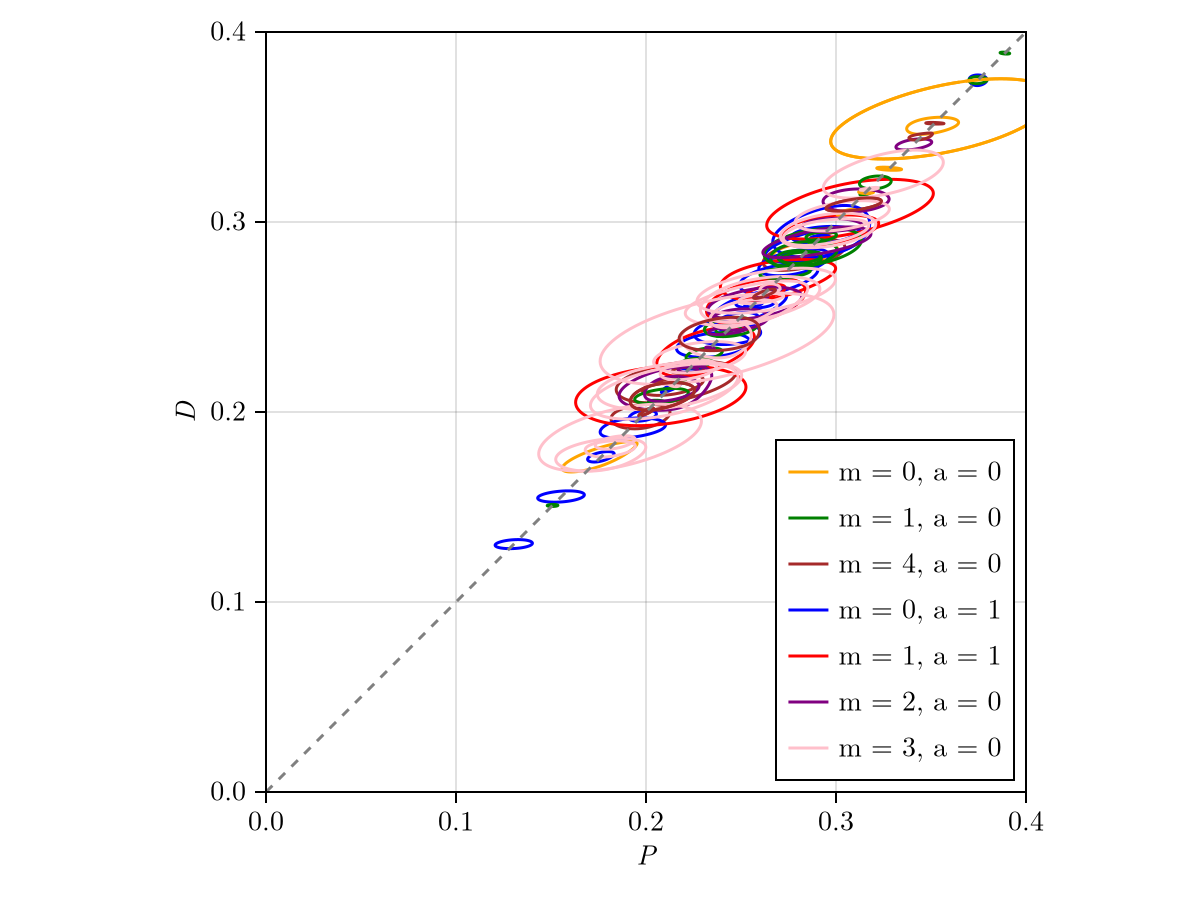}
    \caption{The rate of energy dissipation $D$ against production $P$ for the successfully converged solutions. The colours show the different symmetry transformations relative to which they are periodic orbits. Compare with \cref{fig:dpgeneric}. Two orbits with $D\approx 0.6$ are excluded from this plot.}
    \label{fig:allorbits}
\end{figure}

The synthetic trajectories were used as initial guesses for the convergence algorithm described in \cref{sec:convergence}.
Each was run on a single NVIDIA Tesla V100 GPU with 32 GB of memory for a maximum of 24 hours, and if the relative error was not less than $10^{-6}$ after this time, the guess was discarded.
Of the $2800$ guesses, 597 converged to a relative error of less than $10^{-6}$, though 103 of these were apparently equilibrium or travelling wave solutions and were not considered further.
$285$ `converged' to orbits with period $T>3$ and were not considered further, both because we are most interested in short orbits but also because $M=64$ is likely to be underresolved and thus the results unreliable when $T\gg3$. A further $57$ orbits were detected to be duplicates and therefore discarded.
The remaining 152 orbits were then reconverged at both $(N,M)=(64,128)$ and $(N,M)=(96,64)$ -- it being too memory intensive to increase in both dimensions simultaneously -- to check that the values were not sensitive to the discretisation. 
In total, 111 unique RPOs with $T< 3$ were found for which the reconvergence at the two different increased resolutions matched, strongly indicating true solutions to the governing equations. Forty of these had $T< 2$, and one of these had $T<1$, though this latter had extremely high dissipation and is omitted from \cref{fig:allorbits}. \Cref{fig:conclose,fig:connotclose} show example synthetic orbits and the corresponding converged solutions. In the former case, the solutions are visually very similar to the synthetic guesses, and in the latter they are very different. Most of our successful convergences were between these two extremes, with some clearly related vortices, but with noticeable differences in the details.
\Cref{tab:fullresults} summarizes the results for each of the symmetry classes. As far as we can tell, none of our solutions were previously known, though it is possible that multiple traversals of these short orbits had been detected by previous authors as longer POs.

\begin{figure}
    \centering
    \resizebox{\linewidth}{!}{\input{figures/comparison-close.tex}}
    \caption{Examples of synthetic orbits which converged to visually similar solutions. In each pair, the generated column depicts vorticity field slices of the synthetic trajectory with $T=1.28$ and the converged column slices of the solution of the Navier--Stokes equations. Rows show equal phase $t/T$, using each trajectory's own period. All snapshots share the vorticity scale shown on the right. The bottom panel shows the dissipation $D$ against production $P$ for the synthetic (black) and converged (coloured) trajectories. Example (a), periodic relative to $\mathcal{T}^s \mathcal{R}^a \mathcal{S}^m$ with $m=1$ and $a=0$ converged to a solution with $s=0$ and  $T\approx1.56$. Example (b) has $m=4$, $a=0$, $s\approx-0.02$ and $T\approx2.46$.}
    \label{fig:conclose}
\end{figure}

\begin{figure}
    \centering
    \resizebox{\linewidth}{!}{\input{figures/comparison-different.tex}}
    \caption{Examples of synthetic orbits which converged to very different solutions, as per \cref{fig:conclose}. Example (a), periodic relative to $\mathcal{T}^s \mathcal{R}^a \mathcal{S}^m$ with $m=1$ and $a=0$ converged to a solution with $s=0$ and  $T\approx2.33$. Example (b) has $m=0$, $a=1$, $s=0$ and $T\approx2.86$.}
    \label{fig:connotclose}
\end{figure}

\begin{table}
\centering
\begin{tabular}{*{8}{c}}
$m$ &
$a$ &
Generated &
\makecell{Did not\\converge} &
\makecell{Equilibria/\\TWs} &
\makecell{Too\\long} &
\makecell{Unique\\candidates} &
\makecell{Converged\\RPOs} \\
\hline
0&0&400&345&4&41&9&9\\
1&0&400&285&21&35&32&21\\
2&0&400&344&4&33&17&13\\
3&0&400&311&10&29&31&23\\
4&0&400&287&21&67&22&16\\
0&1&400&276&39&52&28&21\\
1&1&400&355&4&28&13&8\\\hline
&&2800&2203&103&285&152&111\\
\end{tabular}
\caption{The number of solutions converged for each of the $7$ distinct classes of symmetry. More than 20\% of the generated synthetic orbits converged at low resolution, but many of these converged to equilibria or travelling waves, to period orbits with $T>3$, or to duplicates of one another. The unique converged candidates were then reconverged at higher resolution, giving the totals in the final column.}
\label{tab:fullresults}
\end{table}

\subsection{$\mathcal{R}$-invariant solutions}
\label{sec:Rresults}
It was observed by \citet{chandler2013invariant} that the $\mathcal{R}$-invariant subspace seems to be closely visited by chaotic trajectories in the full space. In particular, orbit P1 in table 2 of that work is invariant under rotation by $\pi$ and has period $T\approx5.38$. This orbit has been found numerous times by subsequent authors \citep{parker2022variational, redfern2024dynamically}. This is not an RPO, it is a full periodic orbit: restricting to the $\mathcal{R}$-invariant subspace removes the continuous symmetry.

The diffusion model and convergence method presented in this paper will not naturally find solutions in invariant subspaces, though several examples in the previous section were close to these, likely representing weak symmetry-breaking of solutions from these invariant subspaces. Because of the symmetry-aware design of the diffusion model, to search directly for $\mathcal{R}$-invariant POs we can simply use as input for our diffusion model a noise field which itself is invariant under rotation by $\pi$.
The same construction applies to other invariant subspaces whose defining symmetries are exactly preserved by the discretized network; it does not automatically extend to every symmetry of the continuous equations.

It is important to note that we did not retrain the network to generate these $\mathcal{R}$-invariant POs; we used the same weights as in the previous subsection. This means that the dynamics we generate are not likely to be representative of the dynamics in this invariant subspace. Indeed, we did not even verify that the dynamics are chaotic in this case. Rather, we aim to generate invariant POs which are relevant to the chaotic dynamics of the full space.

As in \cref{sec:fullspaceresults}, we generated 400 synthetic orbits. For each input to the diffusion model we took a $64\times64\times64$ unit Gaussian noise field, which was then rotated by $\pi$ in the spatial dimensions, added to the original and then the result was divided by $\sqrt{2}$ to give an $\mathcal{R}$ invariant noise field matching the same unit standard deviation as the general Gaussian noise fields used to train the model. These noise fields were then subject to backward diffusion in exactly the same way as in \cref{sec:fullspaceresults}, and the resulting synthetic orbits used as initial guesses for our convergence algorithm.

\begin{figure}
    \centering
    \includegraphics[width=0.7\linewidth]{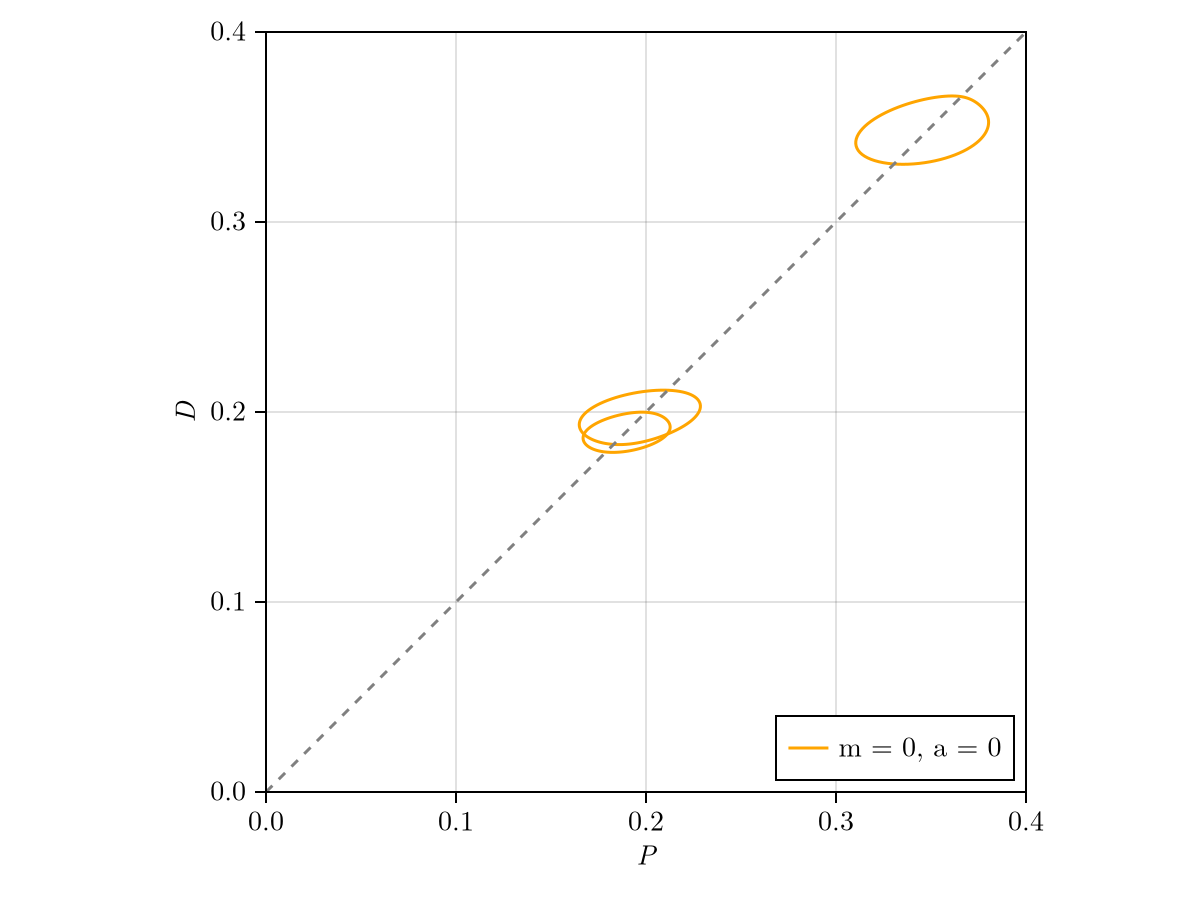}
     \caption{The dissipation $D$ against energy production $P$ for the successfully converged periodic orbits in the $\mathcal{R}$-invariant subspace. One orbit with $D\approx 0.69$, depicted in \cref{fig:rsubcon}, is excluded from this plot. }
    \label{fig:symmorbits}
\end{figure}

Of the 400 generated synthetic trajectories, 91 converged to a residual below our cutoff, but 16 of these were equilibria and thus discarded. Of the remaining 75, all but one converged to have period greater than our cut-off of $T=3$, and this single success converged to an orbit with very high dissipation. When the cut-off was relaxed to $T\leq4$, a further three likely physical periodic orbits were detected. All four of these were successfully reconverged at $(N,M)=(64,128)$ and $(N,M)=(96,64)$.
All four of these had high dissipation. Two examples are shown in \cref{fig:rsubcon}, and the full collection is depicted in \cref{fig:symmorbits}.

\begin{figure}
    \centering
    \resizebox{\linewidth}{!}{\input{figures/comparison-r-invariant.tex}}
    \caption{Successfully converged solutions in the $\mathcal{R}$-invariant subspace, with columns, phase labels and the shared vorticity scale as in \cref{fig:conclose}. (a) This synthetic orbit 
    converged to a PO with period $T\approx 2.17$ and the extremely high dissipation $D\approx0.69$. The converged solution appears to also be invariant under a horizontal shift $\mathcal{T}^\pi$, which was certainly not true for the guess. (b) This synthetic orbit 
    converged to a PO with $T\approx3.90$. The converged result is visually similar to the synthetic orbit, though the dissipation is still somewhat higher.}
    \label{fig:rsubcon}
\end{figure}

\section{Discussion}
\label{sec:discussion}

\subsection{Short-cycle abundance and periodic-orbit descriptions}

Our results resolve the question of whether the small number of previously known very short periodic orbits reflects an intrinsic scarcity of such solutions in two-dimensional Kolmogorov flow. The 40 relative periodic orbits with \(T<2\) identified here, compared with the three in the Edinburgh reference collection \citep{edinburgh2026orbits}, demonstrate that very short cycles are in fact abundant. 

All of the short converged orbits we have found are geometrically very simple, tracing the oval loops in the $P-D$ plane illustrated in \cref{fig:allorbits}. Each executes an elementary recurrent motion, without the successive excursions around other cycles that would identify it as a composite orbit shadowing a sequence of shorter ones. Assuming that most or all of these solutions belong to the chaotic attractor, their geometric simplicity implies that each represents a single symbol in a symbolic dynamics organized around these elementary motions. Thus, the 111 orbits already require a large alphabet, before longer orbits formed by combining these motions are considered. Furthermore, we found relatively few duplicates and did not make a concerted effort to find orbits confined to $D\leq0.15$, suggesting that many more elementary cycles remain to be found.

The implications become clear at the next symbolic order. A two-symbol prime cycle $ab$ follows the elementary motion $a$ and then $b$ before closing. The words $ab$ and $ba$ describe the same periodic itinerary, while $aa$ is a repeat of the one-symbol cycle and is not a new prime cycle. An alphabet of $N$ symbols therefore has $N(N-1)/2$ possible two-symbol prime itineraries before pruning by the dynamics \citep{ChaosBook}. For the 111 elementary cycles already found, this gives 6105 possible pairings. We do not expect every pairing to correspond to an orbit: the allowed transitions constitute the grammar of the symbolic dynamics, and forbidden itineraries must be pruned. Nevertheless, determining this grammar and computing the surviving cycles is itself a substantial task. If the elementary population extends to several hundred or a thousand cycles, as the incomplete coverage of the present search suggests is possible, the corresponding unpruned collection reaches tens or hundreds of thousands of two-symbol itineraries.

This proliferation directly affects the practical implementation of periodic orbit theory. In a cycle expansion, a longer prime cycle is grouped with products of shorter-cycle weights. Shadowing can leave small curvature corrections such as $t_{ab}-t_a t_b$, where $t_p$ denotes a cycle weight incorporating its period, stability and the observable of interest \citep{ChaosBook,lan2010cycle}. To exploit these cancellations systematically, one must identify the admissible composite cycles, converge them, and calculate their stability weights. A large elementary alphabet makes this a major computational undertaking already at symbolic length two, even before higher-order combinations are included. On this basis, we regard a systematic periodic-orbit-theory calculation for this flow as computationally impractical with current methods. 

Nevertheless, statistical reconstructions from empirically weighted sums of period orbits remain useful in this setting. Good estimates of selected observables can be obtained from comparatively few reference trajectories without constructing a complete symbolic truncation \citep{page2024recurrent,redfern2024dynamically,pughe2025computing}. Their success does not resolve the computational burden of identifying and evaluating the elementary and composite cycles required for a systematic cycle expansion.

\subsection{Generative models for exact-solution discovery}

The success of our approach relies on the complementary capabilities of generative modelling and robust numerical refinement. The trajectories produced by the diffusion model are statistically plausible but generally remain poor approximations to exact solutions of the Navier--Stokes equations. Such guesses are difficult to exploit with shooting-based methods, which rely on time integration and require the initial trajectory to return sufficiently close to itself after one period. In contrast, the space-time variational method operates on the entire candidate loop and iteratively deforms the full space-time field toward an exact solution. Its large convergence region therefore allows comparatively inaccurate generated trajectories to serve as useful initial guesses, while the generative model provides a means of exploring the vast space of possible recurrent motions. The two approaches thus address complementary aspects of the discovery problem: generating diverse plausible candidates and robustly converging them to exact solutions.

This complementarity suggests a broader role for generative learning in physics. In physics-informed neural networks, governing-equation residuals enter the learning objective \citep{raissi2019physics}. Recent physics-informed generative models such as FunDiff also incorporate physical constraints through their architecture or loss \citep{wang2026fundiff}. For rotating three-dimensional turbulence, \citet{li2026generative} impose constraints within the diffusion process to synthesize single-time velocity fields. Our model uses physical knowledge through its symmetries and forcing-dependent architecture, but its training objective contains no Navier--Stokes residual. The requirement of actually solving the evolution equations is instead imposed during subsequent numerical refinement and checked independently of the learned model.

The resulting discoveries are interpretable objects in the original dynamical system: invariant solutions specified by their vorticity fields, periods and symmetry transformations. Their vortex motions, stability and relation to turbulent trajectories can be analyzed independently of the network that generated the initial guesses. This provides an inspectable scientific output without implying that the network's internal representation is itself interpretable. Generative learning thus supplies candidates for exploration, while the equations and numerical verification determine which candidates qualify as invariant solutions.

The present results also illustrate that changing how invariant solutions are searched for can alter the resulting physical picture, rather than merely accelerate their discovery. Existing search strategies had revealed only a small number of very short relative periodic orbits, leaving open whether such solutions were intrinsically rare. The generative approach samples candidate recurrent motions in a qualitatively different way and reveals instead a large and diverse short-period population. Generative models may therefore be particularly useful not simply for finding more solutions, but for exploring the invariant-solution landscape through search mechanisms complementary to those based on recurrences observed directly in turbulent simulations. In the present case, this expanded exploration exposes previously unresolved structure in the recurrent dynamics and motivates new questions about the organization of periodic orbits and their role in describing turbulent statistics.

The present study therefore raises two related questions for future work. On the methodological side, it remains to be determined how broadly the combination of generative exploration and robust space-time refinement can be transferred to other nonlinear PDEs and turbulent flows. On the physical side, systematic searches in other systems are needed to establish whether the rich short-period orbit population found here is characteristic of spatially extended chaos or specific to the present flow, and whether the same computational obstruction to systematic cycle expansions persists in those systems. Addressing these questions will test both the generality of the discovery strategy and the emerging picture of how recurrent dynamics are organized in high-dimensional chaotic flows.

\begin{acknowledgments}
\textbf{Funding.} This work was supported by the European Research Council (ERC) under the European Union's Horizon 2020 research and innovation programme (grant no. 865677).

\textbf{Use of artificial intelligence.} ChatGPT and GitHub Copilot were used to assist with writing some Python code snippets for the diffusion model. All Julia research code was human-written. GPT-6.1 Sol was used to clean the research code to produce the publicly available version \url{https://github.com/jeremypparker/Parker_Schneider_Diffusion_Model}. The results of this AI-cleaned code have been verified against the original outputs. ChatGPT was used for iterating this manuscript. The schematic \cref{fig:kernels,fig:conv,fig:unet} were generated by GPT-6 Astra based on initial versions created by the authors. The authors take full responsibility for the results of this work and the content of the manuscript.

\textbf{Declaration of interests.} The authors report no conflict of interest.

\end{acknowledgments}

\begin{appen}
\crefalias{section}{appendix}

\section{Symmetries}
\label{sec:symmetries}
Let $\mathcal{U}$, $\mathcal{X}$, $\mathcal{Y}$ and $\mathcal{C}$ represent, respectively, $y$-shift by $\pi/4$, reflection in $x$, reflection in $y$ and change of sign of $\omega$. Considering these alongside $\mathcal{T}^s$, an $x$-shift by $s$, as the fundamental transformations of the vorticity field $\omega$, we have the commutation relations $\mathcal{U}\mathcal{Y} = \mathcal{Y}\mathcal{U}^{-1}$ and $\mathcal{X}\mathcal{T}^s=\mathcal{T}^{-s}\mathcal{X}$ and all other combinations commute. Also note that $\mathcal{X}$, $\mathcal{Y}$ and $\mathcal{C}$ are their own inverses.
The basic symmetries of the system, under which \cref{eq:vorticityeq} is equivariant, are then the discrete symmetries $\mathcal{S}=\mathcal{C}\mathcal{U}\mathcal{X}$ and $\mathcal{R} = \mathcal{X}\mathcal{Y}$ and the continuous symmetry $\mathcal{T}^s$ for any $s$. (If $2\pi/s$ is a rational number then after raising $\mathcal{T}^s$ to some integer power we recover the identity, so in that sense it is a discrete symmetry, but still belongs to the one-parameter family.)

Using these definitions, we can immediately derive the commutation relations for the symmetries
\begin{equation}\mathcal{R}\mathcal{S} = \mathcal{S}^{-1}\mathcal{R},
\quad\mathcal{S}\mathcal{R} = \mathcal{R}\mathcal{S}^{-1},\quad\mathcal{T}^s\mathcal{R} = \mathcal{R}\mathcal{T}^{-s},\quad\mathcal{T}^s\mathcal{S} = \mathcal{S}\mathcal{T}^{-s}.\end{equation}
The subgroup generated by $\mathcal{R}$ and $\mathcal{S}$ has the presentation
\begin{equation}
    \left<\mathcal{R},\mathcal{S} \,|\, \mathcal{R}^2, \mathcal{S}^8, (\mathcal{R}\mathcal{S})^2\right> \cong D_8,\label{eq:group}
\end{equation}
isomorphic to the dihedral group of order 16.

\subsection{Inverses}

A general symmetry can be written in the form
\begin{align}\begin{split}
    \mathcal{T}^s \mathcal{R}^a \mathcal{S}^m &= \mathcal{T}^s \mathcal{X}^a \mathcal{Y}^a \mathcal{C}^m \mathcal{U}^m \mathcal{X}^m \\
    &= \mathcal{T}^s \mathcal{U}^{(-1)^a m} \mathcal{Y}^a \mathcal{X}^{a+m} \mathcal{C}^m,
\end{split}\end{align}
for $a\in\{0,1\}$, $m\in\{0,\dots,7\}$ and $s\in[0,2\pi)$.
When $a=0$, this has inverse
\begin{align}\begin{split}
    \left(\mathcal{T}^s \mathcal{R}^0 \mathcal{S}^m\right)^{-1} &= \mathcal{S}^{-m} \mathcal{T}^{-s} \\
    &= \mathcal{C}^{-m} \mathcal{U}^{-m} \mathcal{X}^{-m}  \mathcal{T}^{-s} \\
    &= \mathcal{T}^{-(-1)^{-m} s} \mathcal{U}^{-m} \mathcal{X}^{-m} \mathcal{C}^{-m}\\
    &= \mathcal{T}^{(-1)^{1+m}s} \mathcal{R}^0 \mathcal{S}^{-m},
\end{split}\end{align}
and for $a=1$,
\begin{align}\begin{split}
    \left(\mathcal{T}^s \mathcal{R}^1 \mathcal{S}^m\right)^{-1} &= \mathcal{S}^{-m} \mathcal{R} \mathcal{T}^{-s} \\
    &= \mathcal{C}^{-m} \mathcal{U}^{-m} \mathcal{X}^{1-m} \mathcal{Y} \mathcal{T}^{-s} \\
    &= \mathcal{T}^{-(-1)^{1-m} s} \mathcal{U}^{-m} \mathcal{Y}^1 \mathcal{X}^{1+m} \mathcal{C}^{m}\\
    &= \mathcal{T}^{(-1)^{m}s} \mathcal{R}^1 \mathcal{S}^{m},
\end{split}\end{align}
with the exponents of $\mathcal{T}$ and $\mathcal{S}$ being understood modulo $2\pi$ and $8$ respectively.

\subsection{Equivalent solutions}
Suppose we have an initial condition $\omega_0$ giving rise to an RPO relative to $\mathcal{T}^s \mathcal{R}^a \mathcal{S}^m$, i.e.
\begin{equation}
\omega_T = \mathcal{T}^s \mathcal{R}^a \mathcal{S}^m \omega_0.
\end{equation}
Then
\begin{equation}
\mathcal{R}\omega_T = \mathcal{R}\mathcal{T}^s \mathcal{R}^a \mathcal{S}^m \omega_0 = \mathcal{T}^{-s} \mathcal{R}^a \mathcal{S}^{-m} (\mathcal{R}\omega_0),
\end{equation}
which means that any RPOs with $m=5,6,7$ are equivalent to ones with $m=3,2,1$ respectively, after rotating the vorticity field by $\pi$.
Furthermore,
\begin{equation}
\mathcal{S}\omega_T = \mathcal{S}\mathcal{T}^s \mathcal{R}^a \mathcal{S}^m \omega_0 = \begin{cases}\mathcal{T}^{-s} \mathcal{R}^a \mathcal{S}^m (\mathcal{S}\omega_0), &a=0,\\\mathcal{T}^{-s} \mathcal{R}^a \mathcal{S}^{m-2} (\mathcal{S}\omega_0), &a=1,\end{cases}
\end{equation}
so the classes with $a=1$ and $m$ even are all equivalent, likewise for $a=1$ and $m$ odd.
If $a=1$ and $m$ is even, we have
\begin{equation}
\mathcal{T}^{-s/2}\omega_T = \mathcal{T}^{s/2}\mathcal{R}^a \mathcal{S}^m \omega_0 = \mathcal{R}^a \mathcal{S}^m (\mathcal{T}^{-s/2}\omega_0)
\end{equation}
so the orbit is equivalent to one with $s=0$.
If $a=0$ and $m$ is odd, the same result holds.

Consequently, it is sufficient to consider representatives with
\begin{equation}
\begin{alignedat}{3}
a&=0,\qquad &m&\in\{0,2,4\},\qquad &s&\in[0,2\pi),\\
a&=0,      &m&\in\{1,3\},         &s&=0,\\
a&=1,      &m&=0,                 &s&=0,\\
a&=1,      &m&=1,                 &s&\in[0,2\pi).
\end{alignedat}
\label{eq:symmetry-representatives}
\end{equation}
The range of $s$ can be reduced further, but we retain
$[0,2\pi)$ wherever the shift remains unrestricted above.\end{appen}\clearpage

\bibliographystyle{jfm}
\bibliography{references}

@article{raissi2019physics,
  title={Physics-informed neural networks: {A} deep learning framework for solving forward and inverse problems involving nonlinear partial differential equations},
  author={Raissi, Maziar and Perdikaris, Paris and Karniadakis, George E},
  journal={Journal of Computational Physics},
  volume={378},
  pages={686--707},
  year={2019},
  publisher={Elsevier},
  doi={10.1016/j.jcp.2018.10.045}
}

@article{lan2004variational,
  title={Variational method for finding periodic orbits in a general flow},
  author={Lan, Yueheng and Cvitanovi{\'c}, Predrag},
  journal={Physical Review E},
  volume={69},
  number={1},
  pages={016217},
  year={2004},
  publisher={APS}
}

@article{parker2022variational,
  title={Variational methods for finding periodic orbits in the incompressible {Navier--Stokes} equations},
  author={Parker, Jeremy P and Schneider, Tobias M},
  journal={Journal of Fluid Mechanics},
  volume={941},
  pages={A17},
  year={2022},
  publisher={Cambridge University Press},
  doi={10.1017/jfm.2022.299}
}

@article{azimi2022constructing,
  title={Constructing periodic orbits of high-dimensional chaotic systems by an adjoint-based variational method},
  author={Azimi, Sajjad and Ashtari, Omid and Schneider, Tobias M},
  journal={Physical Review E},
  volume={105},
  number={1},
  pages={014217},
  year={2022},
  publisher={APS}
}

@article{parker2023predicting,
  title={Predicting chaotic statistics with unstable invariant tori},
  author={Parker, Jeremy P and Ashtari, Omid and Schneider, Tobias M},
  journal={Chaos: An Interdisciplinary Journal of Nonlinear Science},
  volume={33},
  number={8},
  year={2023},
  publisher={AIP Publishing},
  pages={083111},
  doi={10.1063/5.0143689}
}

@article{page2020searching,
  title={Searching turbulence for periodic orbits with dynamic mode decomposition},
  author={Page, Jacob and Kerswell, Rich R},
  journal={Journal of Fluid Mechanics},
  volume={886},
  pages={A28},
  year={2020},
  publisher={Cambridge University Press}
}

@article{page2024recurrent,
  title={Recurrent flow patterns as a basis for two-dimensional turbulence: {Predicting} statistics from structures},
  author={Page, Jacob and Norgaard, Peter and Brenner, Michael P and Kerswell, Rich R},
  journal={Proceedings of the National Academy of Sciences},
  volume={121},
  number={23},
  pages={e2320007121},
  year={2024},
  publisher={National Acad Sciences}
}

@article{kawahara2001periodic,
  title={Periodic motion embedded in plane {Couette} turbulence: regeneration cycle and burst},
  author={Kawahara, Genta and Kida, Shigeo},
  journal={Journal of Fluid Mechanics},
  volume={449},
  pages={291--300},
  year={2001},
  publisher={Cambridge University Press}
}

@article{kawahara2012significance,
  title={The significance of simple invariant solutions in turbulent flows},
  author={Kawahara, Genta and Uhlmann, Markus and van Veen, Lennaert},
  journal={Annual Review of Fluid Mechanics},
  volume={44},
  pages={203--225},
  year={2012},
  publisher={Annual Reviews},
  doi={10.1146/annurev-fluid-120710-101228}
}

@Book{ChaosBook,
  title =     {Chaos: {Classical} and {Quantum}},
  publisher = {Niels Bohr Inst.},
  year =      {2016},
  author =    {P. Cvitanovi{\'c} and R. Artuso and R. Mainieri and G. Tanner and G. Vattay},
  address =   {Copenhagen},
  url =       {http://ChaosBook.org/},
  note={\url{https://ChaosBook.org/}}
}

@article{viswanath2007recurrent,
  title={Recurrent motions within plane {Couette} turbulence},
  author={Viswanath, Divakar},
  journal={Journal of Fluid Mechanics},
  volume={580},
  pages={339--358},
  year={2007},
  publisher={Cambridge University Press}
}

@article{cvitanovic2010state,
  title={On the state space geometry of the {Kuramoto--Sivashinsky} flow in a periodic domain},
  author={Cvitanovi{\'c}, Predrag and Davidchack, Ruslan L and Siminos, Evangelos},
  journal={SIAM Journal on Applied Dynamical Systems},
  volume={9},
  number={1},
  pages={1--33},
  year={2010},
  publisher={SIAM}
}

@article{chandler2013invariant,
  title={Invariant recurrent solutions embedded in a turbulent two-dimensional {Kolmogorov} flow},
  author={Chandler, Gary J and Kerswell, Rich R},
  journal={Journal of Fluid Mechanics},
  volume={722},
  pages={554--595},
  year={2013},
  publisher={Cambridge University Press}
}

@article{page2024exact,
  title={Exact coherent structures in two-dimensional turbulence identified with convolutional autoencoders},
  author={Page, Jacob and Holey, Joe and Brenner, Michael P and Kerswell, Rich R},
  journal={Journal of Fluid Mechanics},
  volume={991},
  pages={A10},
  year={2024},
  publisher={Cambridge University Press}
}

@article{graham2021exact,
  title={Exact coherent states and the nonlinear dynamics of wall-bounded turbulent flows},
  author={Graham, Michael D and Floryan, Daniel},
  journal={Annual Review of Fluid Mechanics},
  volume={53},
  number={1},
  pages={227--253},
  year={2021},
  publisher={Annual Reviews}
}

@article{beck2024machine,
  title={Data-driven guessing and gluing of unstable periodic orbits},
  author={Beck, Pierre and Parker, Jeremy P and Schneider, Tobias M},
  journal={Physical Review E},
  volume={112},
  number={2},
  pages={024203},
  year={2025},
  publisher={APS}
}

@article{viswanath2001lindstedt,
  title={The {Lindstedt}--{Poincar{\'e}} technique as an algorithm for computing periodic orbits},
  author={Viswanath, Divakar},
  journal={SIAM Review},
  volume={43},
  number={3},
  pages={478--495},
  year={2001},
  publisher={SIAM}
}

@inproceedings{goodfellow2014generative,
  title={Generative adversarial nets},
  author={Goodfellow, Ian J. and Pouget-Abadie, Jean and Mirza, Mehdi and Xu, Bing and Warde-Farley, David and Ozair, Sherjil and Courville, Aaron and Bengio, Yoshua},
  year={2014},
  booktitle={Advances in Neural Information Processing Systems 27},
  publisher={Curran Associates, Inc.}
}

@inproceedings{kingma2013auto,
  title={Auto-encoding variational {Bayes}},
  author={Kingma, Diederik P. and Welling, Max},
  year={2014},
  booktitle={International Conference on Learning Representations},
  url={https://iclr.cc/archive/2014/conference-proceedings/}
}

@inproceedings{sohl2015deep,
  title={Deep unsupervised learning using nonequilibrium thermodynamics},
  author={Sohl-Dickstein, Jascha and Weiss, Eric and Maheswaranathan, Niru and Ganguli, Surya},
  booktitle={Proceedings of the 32nd International Conference on Machine Learning},
  pages={2256--2265},
  year={2015},
  organization={PMLR},
  series={Proceedings of Machine Learning Research},
  volume={37},
  url={https://proceedings.mlr.press/v37/sohl-dickstein15.html}
}

@inproceedings{ho2020denoising,
  title={Denoising diffusion probabilistic models},
  author={Ho, Jonathan and Jain, Ajay and Abbeel, Pieter},
  pages={6840--6851},
  year={2020},
  booktitle={Advances in Neural Information Processing Systems 33},
  publisher={Curran Associates, Inc.}
}

@article{suri2018unstable,
  title={Unstable equilibria and invariant manifolds in quasi-two-dimensional {Kolmogorov-like} flow},
  author={Suri, Balachandra and Tithof, Jeffrey and Grigoriev, Roman O and Schatz, Michael F},
  journal={Physical Review E},
  volume={98},
  number={2},
  pages={023105},
  year={2018},
  publisher={APS}
}

@article{suri2014velocity,
  title={Velocity profile in a two-layer {Kolmogorov-like} flow},
  author={Suri, Balachandra and Tithof, Jeffrey and Mitchell, Jr., Radford and Grigoriev, Roman O and Schatz, Michael F},
  journal={Physics of Fluids},
  volume={26},
  number={5},
  year={2014},
  publisher={AIP Publishing},
  pages={053601},
  doi={10.1063/1.4873417}
}

@article{redfern2024dynamically,
  title={Dynamically relevant recurrent flows obtained via a nonlinear recurrence function from two-dimensional turbulence},
  author={Redfern, Edward M and Lazer, Andrei L and Lucas, Dan},
  journal={Physical Review Fluids},
  volume={9},
  number={12},
  pages={124401},
  year={2024},
  publisher={APS}
}

@article{cleary2025dynamical,
  title={Dynamical relevance of periodic orbits under increasing {Reynolds} number and connections to inviscid dynamics},
  author={Cleary, Andrew and Page, Jacob},
  journal={Journal of Fluid Mechanics},
  volume={1020},
  pages={A52},
  year={2025},
  doi={10.1017/jfm.2025.10673}
}

@inproceedings{song2020denoising,
  title={Denoising diffusion implicit models},
  author={Song, Jiaming and Meng, Chenlin and Ermon, Stefano},
  year={2021},
  booktitle={International Conference on Learning Representations},
  url={https://openreview.net/forum?id=St1giarCHLP}
}

@article{sipper2021neural,
  title={Neural networks with {\`a} la carte selection of activation functions},
  author={Sipper, Moshe},
  journal={SN Computer Science},
  volume={2},
  number={6},
  pages={470},
  year={2021},
  publisher={Springer},
  doi={10.1007/s42979-021-00885-1}
}

@inproceedings{ronneberger2015u,
  title={{U-Net}: {Convolutional} networks for biomedical image segmentation},
  author={Ronneberger, Olaf and Fischer, Philipp and Brox, Thomas},
  booktitle={Medical Image Computing and Computer-Assisted Intervention -- MICCAI 2015},
  pages={234--241},
  year={2015},
  organization={Springer},
  series={Lecture Notes in Computer Science},
  volume={9351},
  doi={10.1007/978-3-319-24574-4_28}
}

@article{platt1991investigation,
  title={An investigation of chaotic {Kolmogorov} flows},
  author={Platt, Nathan and Sirovich, Lawrence and Fitzmaurice, Nessan},
  journal={Physics of Fluids A: Fluid Dynamics},
  volume={3},
  number={4},
  pages={681--696},
  year={1991},
  publisher={American Institute of Physics}
}

@article{racca2023predicting,
  title={Predicting turbulent dynamics with the convolutional autoencoder echo state network},
  author={Racca, Alberto and Doan, Nguyen Anh Khoa and Magri, Luca},
  journal={Journal of Fluid Mechanics},
  volume={975},
  pages={A2},
  year={2023},
  publisher={Cambridge University Press}
}

@article{de2023data,
  title={Data-driven low-dimensional dynamic model of {Kolmogorov} flow},
  author={P{\'e}rez De Jes{\'u}s, Carlos E. and Graham, Michael D},
  journal={Physical Review Fluids},
  volume={8},
  number={4},
  pages={044402},
  year={2023},
  publisher={APS},
  doi={10.1103/PhysRevFluids.8.044402}
}

@article{farazmand2017variational,
  title={A variational approach to probing extreme events in turbulent dynamical systems},
  author={Farazmand, Mohammad and Sapsis, Themistoklis P},
  journal={Science Advances},
  volume={3},
  number={9},
  pages={e1701533},
  year={2017},
  publisher={American Association for the Advancement of Science}
}

@article{lan2010cycle,
  title={Cycle expansions: {From} maps to turbulence},
  author={Lan, Y},
  journal={Communications in Nonlinear Science and Numerical Simulation},
  volume={15},
  number={3},
  pages={502--526},
  year={2010},
  publisher={Elsevier}
}

@article{cvitanovic1995dynamical,
  title={Dynamical averaging in terms of periodic orbits},
  author={Cvitanovi{\'c}, Predrag},
  journal={Physica D: Nonlinear Phenomena},
  volume={83},
  number={1-3},
  pages={109--123},
  year={1995},
  publisher={Elsevier}
}

@inproceedings{vaswani2017attention,
  title={Attention is all you need},
  author={Vaswani, Ashish and Shazeer, Noam and Parmar, Niki and Uszkoreit, Jakob and Jones, Llion and Gomez, Aidan N and Kaiser, {\L}ukasz and Polosukhin, Illia},
  year={2017},
  booktitle={Advances in Neural Information Processing Systems 30},
  publisher={Curran Associates, Inc.}
}

@article{page2021revealing,
  title={Revealing the state space of turbulence using machine learning},
  author={Page, Jacob and Brenner, Michael P and Kerswell, Rich R},
  journal={Physical Review Fluids},
  volume={6},
  number={3},
  pages={034402},
  year={2021},
  publisher={APS}
}

@article{lucas2022stabilization,
  title={Stabilization of exact coherent structures in two-dimensional turbulence using time-delayed feedback},
  author={Lucas, Dan and Yasuda, Tatsuya},
  journal={Physical Review Fluids},
  volume={7},
  number={1},
  pages={014401},
  year={2022},
  publisher={APS}
}

@article{RN40,
   author = {Kaszás, Bálint and Haller, George},
   title = {Joint reduced model for the laminar and chaotic attractors in plane {Couette} flow},
   journal = {Journal of Fluid Mechanics},
   volume = {1023},
   ISSN = {0022-1120},
   DOI = {10.1017/jfm.2025.10795},
   url = {https://dx.doi.org/10.1017/jfm.2025.10795},
   year = {2025},
   type = {Journal Article},
  pages={A1}
}

@article{pughe2025computing,
  title={Computing chaotic time-averages from few periodic or non-periodic orbits},
  author={Pughe-Sanford, Joshua L. and Quinn, Sam and Balabanski, Teodor and Grigoriev, Roman O.},
  journal={Chaos: An Interdisciplinary Journal of Nonlinear Science},
  volume={35},
  number={6},
  pages={063133},
  year={2025},
  doi={10.1063/5.0264212}
}

@misc{liu2026relative,
  title={Relative periodic orbits in spatially extended {Kolmogorov} turbulence: from global to localized recurrence},
  author={Liu, Yaning and Constante-Amores, C. Ricardo},
  eprint={2609.28981},
  year={2026},
  doi={10.48550/arXiv.2609.28981}
}

@article{wang2026fundiff,
  title={{FunDiff}: diffusion models over function spaces for physics-informed generative modeling},
  author={Wang, Sifan and Dou, Zehao and Shan, Siming and Liu, Tong-Rui and Lu, Lu},
  journal={Nature Communications},
  volume={17},
  pages={5749},
  year={2026},
  doi={10.1038/s41467-026-72292-0}
}

@article{li2026generative,
  title={A generative diffusion framework for physically consistent {3D} turbulence},
  author={Li, Tianyi and Buzzicotti, Michele and Bonaccorso, Fabio and Biferale, Luca},
  journal={Nature Communications},
  year={2026},
  doi={10.1038/s41467-026-78265-7}
}

@misc{edinburgh2026orbits,
  author={{invariant-sets.org}},
  title={{Kolmogorov RPOs Re=40 (Edinburgh)}},
  year={2026},
  howpublished={\url{https://invariant-sets.org/}},
  note={Accessed 15 October 2026}
}

@article{ashtari2026computing,
  title={Computing Simple Invariant Solutions in Chaotic Fluid Flows},
  author={Ashtari, Omid and Tuckerman, Laurette S and Schneider, Tobias M},
  journal={Annual Review of Fluid Mechanics},
  volume={59},
  year={2026},
  publisher={Annual Reviews}
}

@article{lucas2015recurrent,
  title={Recurrent flow analysis in spatiotemporally chaotic 2-dimensional {Kolmogorov} flow},
  author={Lucas, Dan and Kerswell, Rich R},
  journal={Physics of Fluids},
  volume={27},
  number={4},
  year={2015},
  publisher={AIP Publishing}
}

@article{beck2026identifying,
  title={Identifying recurrent flows in high-dimensional dissipative chaos from low-dimensional embeddings},
  author={Beck, Pierre and Schneider, Tobias M},
  journal={arXiv preprint arXiv:2601.01590},
  year={2026}
}

\end{document}